\documentclass[11pt]{article}

\usepackage{hyperref}
\usepackage[usenames,dvipsnames]{xcolor}
\usepackage{graphicx}
\usepackage{authblk}
\usepackage{amsmath,amssymb,amsfonts}
\usepackage{cite}

\usepackage[letterpaper, total={16cm, 21cm}]{geometry}

\makeatletter
\g@addto@macro\bfseries{\boldmath}
\let\Hy@backout\@gobble
\makeatother

\let\oldthebibliography\thebibliography
\renewcommand\thebibliography[1]{%
  \oldthebibliography{#1}%
  \setlength{\itemsep}{0pt}%
}

\catcode`\@=11
\renewcommand\paragraph{\@startsection{paragraph}{4}{\z@}%
                                    {1.5ex \@plus.2ex \@minus.2ex}%
                                    {-1em}%
                                    {\normalfont\normalsize\bfseries}}
\catcode`\@=12

\def\OMIT#1{{}}
\def\lqcd{\Lambda_{\rm QCD}}
\newcommand{\Bbar}{\,\overline{\!B}{}}
\newcommand{\Dbar}{\,\overline{\!D}{}}
\newcommand{\Kbar}{\,\overline{\!K}{}}
\def\B0bar{\Bbar{}^0}
\def\D0bar{\Dbar{}^0}
\def\K0bar{\Kbar{}^0}

\def\GeV{{\rm GeV}}

\usepackage{lineno}
\title{
Theory Techniques for Precision Physics \\ Snowmass 2021 TF06 Topical Group Report
\\[4pt]
\begin{centering}
{\large {\bf Convenors:}  Radja Boughezal$^1$, Zoltan Ligeti$^2$}
\end{centering}
}

\author[3]{Wolfgang Altmannshofer}
\affil[1]{High Energy Physics Division, Argonne National Laboratory, Argonne, IL 60439, USA}
\affil[2]{Lawrence Berkeley National Laboratory, University of California, Berkeley, CA 94720, USA}
\affil[3]{Santa Cruz Institute for Particle Physics, Santa Cruz, CA 95064, USA}

\author[4]{Supratim~Das Bakshi}
\affil[4]{Departamento de F\'isica Te\'orica y del Cosmos, Universidad de Granada, %Campus de Fuentenueva, 
E--18071 Granada, Spain}

\author[5]{Fabrizio Caola}
\affil[5]{Rudolf Peierls Centre for Theoretical Physics, and Wadham College, Oxford OX1 3PN, UK}        

\author[4]{Mikael Chala}
\author[4]{Alvaro D\'iaz-Carmona}
\author[6]{Wen~Chen}
\affil[6]{School of Physics, Zhejiang University, Hangzhou, Zhejiang 310027, China}

\author[7,8]{Neda Darvishi}
\affil[7]{Institute of Theoretical Physics, Chinese Academy of Sciences, Beijing 100190, China}
\affil[8]{Department of Physics and Astronomy, Michigan State University, East Lansing, MI 48824, USA}
\author[9]{Brian Henning}
\affil[9]{Institute of Physics, \'Ecole Polytechnique F\'ed\'erale de Lausanne (EPFL), CH-1015 Lausanne, Switzerland}

\author[10]{Sebastian~Jaskiewicz}
\affil[10]{Institute for Particle Physics Phenomenology, Durham University, Durham DH1 3LE, UK}

\author[11]{Teppei~Kitahara}
\affil[11]{Institute for Advanced Research \& Kobayashi-Maskawa Institute, Nagoya University, Nagoya 464-8602, Japan}

\author[12]{Hao-Lin Li}
\affil[12]{Centre for Cosmology, Particle Physics and Phenomenology (CP3), Universite Catholique de Louvain}

\author[13]{Xiaohui Liu}
\affil[13]{Center of Advanced Quantum Studies, Department of Physics, Beijing Normal University, Beijing 100875}

\author[14]{Adam Martin}
\affil[14]{Department of Physics, University of Notre Dame, Notre Dame, IN, 46556, USA}

\author[10]{M.~R.~Masouminia}
\author[15]{Tom Melia}
\affil[15]{Kavli IPMU (WPI), UTIAS, The University of Tokyo, Kashiwa, Chiba 277-8583, Japan}

\author[16]{Emanuele Mereghetti}
\affil[16]{Theory Division, Los Alamos National Laboratory, Los Alamos, NM 87544, USA}

\author[17]{Bernhard Mistlberger}
\affil[17]{SLAC National Accelerator Laboratory, Stanford University, Stanford, CA 94039, USA}

\author[18]{Christopher Murphy}
\affil[18]{Homepoint, Ann Arbor, MI 48105, USA}

\author[1,19]{Frank Petriello}
\affil[19]{Department of Physics and Astronomy, Northwestern University, Evanston, IL 60208, USA}

\author[20]{Davison Soper}
\affil[20]{Institute for Fundamental Science, University of Oregon, Eugene, OR 97403-5203, USA}

\author[21]{George Sterman}
\affil[21]{C.N. Yang Institute for Theoretical Physics and Department of Physics and Astronomy\\
Stony Brook University, Stony Brook NY 11794-3840 USA}

\author[22]{Robert~Szafron}
\affil[22]{Department of Physics, Brookhaven National Laboratory, Upton, NY 11973, USA}

\author[23]{Leonardo Vernazza}
\affil[23]{INFN sezione di Torino}

\author[16]{Gherardo~Vita}
\author[24]{Stefan Weinzierl}
\affil[24]{Johannes Gutenberg-Universit{\"a}t Mainz, D-55099 Mainz, Germany}

\author[6,25]{Jiang-Hao~Yu}
\affil[25]{School of Physical Sciences, University of Chinese Academy of Sciences, Beijing, China}

\author[26]{Jure~Zupan}
\affil[26]{Department of Physics, University of Cincinnati, Cincinnati, Ohio 45221, USA}

\date{}                     %% if you don't need date to appear

\newcommand\snowmass{\vspace*{-14pt}
\begin{center}\rule[-0.2in]{\hsize}{0.01in}\\* \rule{\hsize}{0.01in}\\[-2pt]
\vskip 0.1in 
Submitted to the  Proceedings of the US Community Study\\
on the Future of Particle Physics (Snowmass 2021)\\[-2pt] 
\rule{\hsize}{0.01in}\\* \rule[+0.2in]{\hsize}{0.01in} \end{center}}

\begin{document}

\maketitle

%\vfill
%Change next line for formatting of Snowmass banner
\vspace*{-1.35cm}
\snowmass
\vspace*{-16pt}
\newpage

\begin{abstract}

The wealth of experimental data collected at laboratory experiments suggests that there is some scale separation between the Standard Model (SM) and phenomena beyond the SM (BSM). New phenomena can manifest itself as small corrections to SM predictions, or as signals in processes where the SM predictions vanish or are exceedingly small. This makes precise calculations of the SM expectations essential, in order to maximize the sensitivity of current and forthcoming experiments to BSM physics. This topical group report highlights some past and forthcoming theory developments critical for maximizing the sensitivity of the experimental program to understanding Nature at the shortest distances.

\end{abstract}

%\newcommand\snowmass{\begin{center}\rule[-0.2in]{\hsize}{0.01in}\\\rule{\hsize}{0.01in}\\
%\vskip 0.1in Submitted to the  Proceedings of the US Community Study\\ 
%on the Future of Particle Physics (Snowmass 2021)\\ 
%\rule{\hsize}{0.01in}\\\rule[+0.2in]{\hsize}{0.01in} \end{center}}

%\snowmass

\vspace{0.2 cm}

\section{Executive summary}
Theoretical techniques for precision physics are the backbone of a
successful program in particle physics. When combined with advances at
current colliders such as the LHC or Belle II, and with planned future
experiments, they allow determinations of fundamental parameters of the Standard Model (SM) to unprecedented precision and probe beyond SM (BSM) physics to very high scales. Potential dark matter candidates, explanations of the hierarchy puzzle, and many other discoveries could be made, even if the mechanisms underlying them are at the tens of TeV scale or even beyond.  

To reach the precision level needed to maximize the discovery
potential for new physics, a plethora of challenges in theoretical
physics must be overcome. Calculations beyond leading order in the
perturbative expansions of QCD and the electroweak theory are
needed.  These require the investigation of new mathematical
structures that describe the multi-loop integrals with several mass
scales that occur when considering the Higgs, $W$ and $Z$ bosons, and
top and bottom quarks.  A common theme between collider and flavor
physics is the need for rigorous predictions that combine fixed-order
perturbative calculations, resummations, and nonperturbative
ingredients. Subtraction schemes must be extended to handle the cancellation of
infrared singularities to the N$^3$LO level in perturbation theory,
the order that is expected to become the precision standard for
colliders after the LHC. Resummation of large logarithms involving kinematic or phase space
parameters requires understanding effective field theories (EFTs) at
subleading order in the power expansions. The parton distribution functions for both nucleons and $B$ mesons must be determined to a precision that matches calculations of the parton-level cross sections.
%(both for pdfs in a proton and in a $B$ meson).
Given that current experimental data are consistent with the SM except
for a few anomalies under intense investigation, the understanding of experimental results will inevitably include interpreting them as constraints in the framework of the SM effective field theory (SMEFT). Precision in the SMEFT expansion is needed to properly derive bounds on new physics, and investigations into higher-order corrections in both the loop and EFT expansions are ongoing.  
Improvements in Monte Carlo event generators and parton showers are also critical to maximize the impact of high-precision perturbative calculations, and provide a closer description of experimental data in terms of the hadronic degrees of freedom measured in experiments. 

This topical group addresses the challenges that arise whenever
precision theoretical predictions are required. Its focus lies close to
experiment, since many questions of precision are driven by
experimental opportunities, and is also close to more formal theory and mathematics
due to the issues confronted when going to new orders of
perturbation theory or extending EFTs to new domains. During the
Snowmass process the TF06 working group investigated the current
state-of-the-art in several fields, including higher-order corrections
in both fixed-order and resummed perturbation theory, the
incorporation of a more faithful description of QCD into parton shower
simulations, and calculations within the SMEFT beyond the leading
dimension-6 order. Several whitepapers summarized the status of critical areas relevant to this working group and outlined where future advances are needed. In this report the main activities and findings of the TF06 working group are presented, with attempts to put each topic in the broader context of how they are important to the high energy physics community.

%%%%%
%%%%%%%%%%%%%%%%%%%%%%%%%%%%%%%%%%%%%%%%%%%%%%%%%%%%%%%%%%%%
%\newpage
\section{Precision collider phenomenology}
\subsection{Introduction}

More than 150 inverse femtobarns of integrated luminosity have been delivered by the LHC to the ATLAS and CMS experiments. A high luminosity run of the LHC is expected to deliver integrated luminosity reaching inverse attobarns, as are several future collider experiments under discussion. Detectors at these experiments excel at reconstruction, capable of distinguishing signatures overwhelmed by background such as Higgs boson decays to photons and muons. The systematic and statistical errors at the LHC and other modern experiments now often reach the percent level or below thanks to these experimental advances. With such exquisite data coming now, and with future machines such as an electron-ion collider expected to continue this trend, probes of new physics to very high scales become possible. Potential explanations for puzzling aspects of the SM can be found even if the mechanisms responsible are at the tens of TeV level and possibly even higher.  

This experimental data challenges the theoretical community to predict observables within and beyond the Standard Model (SM) precisely enough to maximize the discovery potential of new physics. The technical difficulties associated with calculations to the N$^3$LO level in QCD in the presence of multiple mass scales, which lead to new mathematical structures such as iterated elliptic integrals, must be confronted. Resummation of large logarithms, including for novel observables such as jet substructure measurements, will require improved understanding of effective field theory (EFT) beyond the usual leading-power approximation. Extractions of parton distributions will need to be upgraded to the  N$^3$LO level in order to match the expected precision of partonic scattering cross sections. Precision calculations within the SM effective field theory (SMEFT) framework will inevitably be part of the future HEP program given that no new particles beyond the SM have yet been discovered. Even in the case of discovery the interpretation of the new states within the SMEFT can help guide future experimental searches by indicating which measurements can help unravel the new state’s identity. Improvements in Monte Carlo event generators and parton showers are also critical to fully exploit the impact of high precision perturbative calculations and provide a closer realization of the actual events in terms of the hadronic degrees of freedom measured in experiments.  

During the Snowmass process the TF06 working group studied numerous challenges that arise in the process of deriving precise predictions for high-$p_T$ physics. These include the path to  N$^3$LO calculations in QCD beyond $2 \to 1$ processes and the requited technical advances, the study of factorization theorems beyond the leading power, improvements in parton showers, and studies at the dimension-8 level in the SMEFT. The work of TF06 is incorporated into several whitepapers that explain these issues, as well as describe the current progress toward their solutions and what will be needed by the future HEP program.  The main activities and findings of the precision collider phenomenology thrust are summarized below, drawing upon the white papers prepared for the Snowmass process.

\subsection{Predictions at NNLO and beyond}

The experimental precision for a host of benchmark processes is approaching the few-percent level. For some measurements in particular clean channels such as Drell-Yan, the claimed experimental uncertainties are below one percent. Such a precision imposes enormous demands on theory. A wide variety of effects must be brought under control to claim a theoretical precision at this level, ranging from exquisite control over parton distribution functions~\cite{Amoroso:2022eow}to the possibility of non-perturbative corrections. One critical component of this program is the calculation of the perturbative hard scattering cross sections to the requisite order in perturbation theory. A famous example of the importance of such higher-order calculations is the cross section for inclusive Higgs boson production at the LHC, for which the higher-order QCD corrections increase the result by nearly a factor of three.

The current standard for cross section predictions at the LHC is next-to-next-to-leading order (NNLO) in perturbative QCD for $2 \to 2$ scattering processes without internal mass scales in the contributing loops. For a host of processes the N$^3$LO corrections in perturbative QCD are available~\cite{Anastasiou:2016cez,Dreyer:2016oyx,Duhr:2019kwi,Chen:2019lzz,Duhr:2021vwj,Duhr:2020seh,Duhr:2020sdp}. The challenge facing the theory community is to extend these computations of hard scattering cross sections at N$^3$LO to include jet processes and more differential observables. Numerous issues must be addressed to achieve this challenge: the computation of the relevant three-loop integrals, understanding the basis of functions needed to describe these corrections, the extension of infrared subtraction schemes to handle triply-unresolved limits, to name only a few. Together with this challenge will also come enormous progress in our understanding of quantum field theory. New understanding of amplitudes, a systematic study of elliptic functions, and novel uses of effective field theory to facilitate such computations have already resulted from this quest.

A roadmap highlighting the current status of the field, and discussing what future progress is needed, was written as part of TF06 working group activities~\cite{Caola:2022ayt}. A brief description of the highlights is given below. The reader is referred to the white paper~\cite{Caola:2022ayt} for more details and complete lists of references.
\begin{itemize}

\item Detailed studies of the renormalization and factorization scale dependences of the N$^3$LO Higgs boson and Drell-Yan cross sections have been performed. One hope of this study is to gain intuition more generally into the size and impact of scale variations at this order in perturbation theory, since these are often used to estimate uncalculated higher-order corrections. An interesting result of this study in the Drell-Yan neutral-current cross sections is that the NNLO scale variation band does not contain the N$^3$LO central value for a range of invariant masses, due to an intricate interplay between the contributing partonic channels~\cite{Duhr:2021vwj,Duhr:2020seh,Duhr:2020sdp}. This behavior is not observed for inclusive Higgs boson production~\cite{Anastasiou:2016cez,Mistlberger:2018etf}.

\item One is inexorably led to the issue of masses in loops when calculating predictions at higher orders due to the presence of numerous mass scales in the SM, most notable the $W$, $Z$, Higgs and top quark masses at the LHC. Even very simple two-loop self-energy diagrams with masses exhibit complicated functional dependences no longer described by the multiple polylogarithms present in the massless case. The  elliptic integrals that appear in massive loop diagrams have received significant attention recently~\cite{Duhr:2019rrs,Weinzierl:2020fyx}, and algorithms for their efficient numerical implementation have been proposed~\cite{Walden:2020odh}. The functional dependences appearing at higher loop order has been reviewed extensively in the following Snowmass whitepaper~\cite{Bourjaily:2022bwx}.

\item Each of the individual components of a higher-order calculation is separately divergent due to infrared singularities (IR). In virtual corrections the IR singularities are always manifest, but in real-radiation corrections they only appear explicitly  after integrating over QCD radiation. Since one is usually interested in fully exclusive results, a method to extract the IR singularities before integration is required. Techniques used in LHC applications for processes with final-state jets included antennae subtraction~\cite{Gehrmann-DeRidder:2005btv,Currie:2013dwa}, sector-improved residue subtraction~\cite{Czakon:2010td,Boughezal:2013uia,Boughezal:2015dra,Caola:2017dug}, and $N$-jettiness subtraction~\cite{Boughezal:2015dva,Gaunt:2015pea}. Other important techniques with numerous important LHC applications include $q_T$-subtraction~\cite{Catani:2007vq} and projection-to-Born~\cite{Cacciari:2015jma}. Significant effort has been devoted to extending these techniques to the N$^3$LO level, and in the case of $q_T$ and $N$-jettiness subtractions many of the ingredients are now known to the required level~\cite{Ebert:2020yqt,Ebert:2020unb,Chen:2022cvz,Baranowski:2022khd}.

\item A wide variety of other issues must be confronted when attempting to push the precision frontier to N$^3$LO, ranging from practical issues such as getting codes to run within the available computational resources~\cite{Cordero:2022gsh}, to the possibility of collinear factorization violations~\cite{Catani:2011st} and enhanced power corrections for observables not inclusive over QCD radiation~\cite{Caola:2021kzt}. These and other issues are surveyed in a TF06 white paper~\cite{Caola:2022ayt}. In particular, it is important to investigate whether the standard collinear factorization approach still applies at this order in perturbation theory. The complication in establishing collinear factorization for hadron collider processes is to show the cancellation of phases arising from soft gluon exchange between the different hard scattering directions. Despite the complications surveyed in~\cite{Caola:2022ayt}, it is expected that for a host of hadron collider processes (annihilation into electroweak particles, inclusive jet production, and heavy-quark production), the cancellation of these contributions does occur, and therefore that the standard collinear factorization does hold~\cite{Sterman:2022gyf}. A more intricate issue that arises is the non-cancellation of these phases in the presence of disparate energy cuts in different regions of the final-state phase space. Such cuts can lead to ``super-leading” logarithmic enhancements and can complicate the standard factorization picture~\cite{Sterman:2022gyf}.

\end{itemize}

Achieving N$^3$LO precision for collider observables requires a determination of the PDFs at the same order. Already the estimated 
uncertainty coming from missing N$^3$LO effects from PDFs is a significant component of the Higgs cross section error budget~\cite{Anastasiou:2016cez}. The progress needed to advance PDFs to the requisite level, including a summary of what pieces of the full N$^3$LO DGLAP evolution are still missing, was extensively reviewed in the following Snowmass whitepaper~\cite{Amoroso:2022eow}.

\subsection{Resummation for future colliders}

Computations of the hard scattering cross section at fixed orders in perturbation theory are sufficient for many applications, but not for all situations.
In multi-scale problems, when one mass scale is very different than the others, large logarithms of 
the scale ratios can appear. Denoting these scales generically as $Q_B$ and $Q_S$ with $Q_S \ll Q_B$, if $\alpha_s {\rm ln}^2(Q_B/Q_S) \sim 1$, then fixed-order perturbation theory doesn't converge. To get a reliable prediction one must resum these large logarithms to all orders. These logarithms can be thought of as an incomplete cancellation between real and virtual corrections. In virtual corrections one integrates over all loop momenta. The real radiation phase space can be restricted by experimental cuts. We can schematically write this situation as
\begin{equation}
\frac{1}{\epsilon} + \int_0^{Q_S} \frac{dQ}{Q} \left( \frac{Q}{Q_B}\right)^{-\epsilon} = {\rm ln}\frac{Q_B}{Q_S}+{\cal O}(\epsilon)
\end{equation}
where $\epsilon$ is the dimensional-regularization parameter that controls infrared divergences. The first term arises from the virtual corrections, while the second term denotes the integration of the real-emission corrections over some final-state phase space. If the phase space is severely restricted so that $Q_S \ll Q_B$, this logarithm becomes large and must be resummed.

There are numerous examples in particle physics where the above situation occurs, most notably in the threshold limit where little energy is available for emission into partonic radiation~\cite{Sterman:1986aj,Catani:1989ne}, or in the low transverse momentum limit where all additional radiation is restricted to be collinear to the beam direction~\cite{Collins:1981uk,Collins:1984kg,Bozzi:2003jy}. These classic examples of all-orders results in QCD have been reformulated in the recent past using the language of soft-collinear effective field theory (SCET)~\cite{Bauer:2000ew,Bauer:2000yr, Bauer:2001yt,Bauer:2002nz}. Other recent applications of these techniques include the small-$x$ limit of QCD and the study of jet substructure. Until recently most applications considered only the leading-power term in the small ratio $Q_S/Q_B$. Terms suppressed by powers of this ratio were dropped. There has recently been significant interest in understanding these sub-leading power terms. This interest is driven by both phenomenological applications, led by the increased precision of the LHC and the expectation of a future electron ion collider, and also by intrinsic theoretical interest in understanding the structure of the resulting factorization theorems at subleading power. Recent results in this field were surveyed as part of the TF06 activities and summarized in the white paper~\cite{vanBeekveld:2022blq}. The reader is referred to the white paper~\cite{vanBeekveld:2022blq} for more details and complete lists of references.

\begin{itemize}

\item The study of threshold corrections beyond the leading-power limit has been considered recently within both a direct diagrammatic approach~\cite{Bahjat-Abbas:2019fqa,vanBeekveld:2021hhv} and using SCET~\cite{Beneke:2017ztn,Feige:2017zci,Beneke:2018rbh}. It was shown to be possible to resum the next-to-leading power (NLP) corrections to leading-logarithmic (LL) accuracy for color-singlet production processes such as Drell-Yan or Higgs production as well as deep inelastic scattering. The numerical impact of the NLP-LL resummation for 
these processes is approximately the same as the NNLL resummation to the leading-power terms, indicating that these corrections are relevant for LHC phenomenology~\cite{Beneke:2019mua,vanBeekveld:2021hhv}. 
 
\item Another interesting example where resummation in NLP terms is necessary is the case of mass-suppressed amplitudes, such as the $b$-quark mediated contribution to gluon-fusion Higgs production. In this case the amplitude goes like $ (\alpha_s/\pi) (m_b^2/m_H^2) {\rm ln}^2(m_H^2/m_b^2)$. The expansion parameter in QCD is $ (\alpha_s/\pi)  {\rm ln}^2(m_H^2/m_b^2) \sim 1$. Even though the amplitude is suppressed by the ratio $m_b^2/m_H^2$, as the Higgs program enters the precision phase this is becoming a limiting uncertainty in the theoretical prediction. The leading-logarithmic resummation of these corrections has been achieved~\cite{Melnikov:2016emg,Liu:2018czl}, resulting in a factor of two reduction of the uncertainty estimate coming from these terms. A SCET analysis of this process was also performed, allowing for a NLL resummation of mass-suppressed effects~\cite{Liu:2019oav,Liu:2020tzd}.

\item Jet substructure techniques have been applied to both SM measurements and beyond-the-SM searches over the past decade, and a host of jet grooming techniques have been developed to remove soft radiation effects that are difficult to account for from first principles in QCD (for reviews please see~\cite{Kogler:2018hem,Marzani:2019hun}). More understanding will be needed in the next phase of the LHC, where detectors will be optimized for the first time to perform precision jet substructure measurements, and at an EIC, where precision measurements of jets across a wide range of energies is expected to be possible. Many novel applications of jet substructure at an EIC, including its use to unravel transverse-momentum distributions and nuclear effects in electron-nucleus collisions, have been proposed. To make full use of these opportunities higher accuracy at both leading and sub-leading power will eventually be required.

\item The small Bjorken-$x$ limit of QCD gives insight into the behavior of gluons within nucleons. In this small-$x$ region, the gluon density grows dramatically and enters the nonlinear regime. The color-glass condensate effective theory (CGC) properly resums logarithms of $x$ that appear in this region~\cite{McLerran:1993ni,McLerran:1993ka}. Higher-order cross sections in this effective theory are notoriously difficult to compute, with negative results appearing in the physical regions of phase space. An approach to solve these issues by coupling SCET to the CGC formalism was recently developed~\cite{Liu:2020mpy}, which will be important when the gluon saturation regime is more thoroughly studied at the EIC.

\end{itemize}

\subsection{The need for precision PDFs}

All hadron collider predictions require understanding of the parton distribution functions that describe how to take a parton of a given momentum fraction from a hadron. Specializing to the collinear parton distribution functions most commonly needed for high-energy collider physics, the PDFs have functional dependence on both the energy scale $\mu$ characteristic of the process under consideration, and on the parton momentum fraction parameterized in terms of the Bjorken momentum fraction $x$. The functional dependence of the PDFs on $\mu$ is perturbative, and is governed by the DGLAP evolution equations. The dependence on $x$ is non-perturbative. Although there has been significant recent activity in attempting to calculate PDFs using lattice techniques~\cite{Constantinou:2020hdm}, for the practical purpose of predicting collider physics cross sections they are extracted from experimental data. This leads to the following three issues that must be addressed in order to have PDFs computed to the needed level for LHC and future collider predictions: the perturbative DGLAP evolution of the PDFs must be calculated to match the precision of the hard scattering cross section; the hard scattering cross sections from which the PDFs are extracted from data must be known to NNLO or N$^3$LO depending on the desired accuracy; the experimental data used in the extraction must have small enough uncertainties to match the above theoretical uncertainties, and must also be sufficiently broad enough to fix the functional dependences of all PDFs on $x$.

The status of PDFs for collider physics, both the current status and desired future improvements, was addressed during the Snowmass process as a joint effort of both TF06 and the energy frontier~\cite{Amoroso:2022eow}. Several findings that address the precision issues raised in the previous paragraph are discussed below.
\begin{itemize}

\item The DGLAP evolution of PDFs is currently known to the NNLO level. To match the desired N$^3$LO level for hard scattering cross sections the calculation of the evolution kernels must be pushed to one order higher. This is currently under active investigation~\cite{Moch:2017uml,Moch:2018wjh,Moch:2021qrk}, and in fact the evolution of the flavor-singlet PDF combination at this order is now possible~\cite{Blumlein:2006be,Blumlein:2021lmf}. The incorporation of heavy quarks into this framework  through the requisite matching calculations, needed to properly evolve the PDFs through mass thresholds where heavy flavors become active, is almost complete to N$^3$LO~\cite{Blumlein:2022ndg}.

\item All modern PDF sets come with uncertainty estimates. These uncertainties account only for the propagation of experimental errors from the data sets used, and do not account for any theoretical uncertainties on the hard scattering cross sections that enter the global fits. There has been recent attempts to incorporate theoretical uncertainties from scale variations into PDF fits by dividing the processes that enter into categories based on their underlying structure and assuming correlated errors for processes with similar structures, allowing a theory covariance matrix to be formed~\cite{NNPDF:2019vjt,NNPDF:2019ubu}. The initial findings indicate NLO PDF determinations are not shifted much by theory uncertainties, but more detailed investigations are expected in future global fits. 

\item Much of our current detailed understanding of PDFs comes from the deep-inelastic scattering experiments at HERA. In the coming decade new DIS data from an electron-ion collider (EIC) is expected, with integrated luminosities reaching an order of magnitude higher than HERA, and with the possibility of polarizing both electron and proton beams~\cite{AbdulKhalek:2021gbh}. The EIC is expected to cover larger-$x$ and smaller-$Q^2$ values than HERA, providing new probes of higher-twist effects that could affect PDF determinations. The anticipated statistical and systematic errors are expected to reach the percent-level or lower. Initial simulations indicate that uncertainties in the valence quark sector could decrease up to 80\% at high-$x$, while the small-$x$ sea quark region could see uncertainty reductions up to 50\%~\cite{Khalek:2021ulf,AbdulKhalek:2021gbh}.

\end{itemize}

\subsection{Future prospects for parton showers}

Both fixed-order and resummed predictions describe collider observables that are inclusive, or differential in a few variables. They are formulated in terms of partonic degrees of freedom. Parton-shower event generators are needed to provide a closer realization of the actual events in terms of hadronic degrees of freedom measured in experiment. Parton showers use the infrared properties of QCD to generate multiple emissions of softer partons starting from a given hard process. At scales $\mu \sim \Lambda_{QCD}$ these partons are combined into hadrons using string or cluster models. Parton-shower event generators such as HERWIG~\cite{Bahr:2008pv}, PYTHIA~\cite{Sjostrand:2014zea} and SHERPA~\cite{Gleisberg:2008ta} are heavily used by experimental collaborations in their analyses and form an indispensable tool for understanding events at high-energy colliders.

There has been numerous theoretical improvements in parton showers over the past years, resulting in programs more faithful to the underlying QCD theory. Matching parton showers to higher-order fixed-order QCD calculations is now available for color-singlet production processes at NNLO, and for arbitrary processes at NLO. Merging of processes with differing numbers of final-state jets is now available at NLO. Algorithms to go beyond the leading-color approximation, as well as to include full spin correlations in order to properly describe azimuthal distributions, are becoming available. One of the advantages of parton-shower simulations is that they inherently resum the singular emissions of QCD, and therefore do not suffer from divergences at kinematic endpoints that often plague fixed-order results, while simultaneously being more flexible than the bespoke resummation calculations described in the previous section. There is an ongoing effort to quantify to exactly what order different parton shower algorithms provide resummation for different observables. At future very high energy colliders it is expected that multiple emissions of electroweak gauge bosons will become important, and there is ongoing work to incorporate electroweak radiation into existing showers.

Parton showers were reviewed as part of the Snowmass process within both the TF06 working group~\cite{Darvishi:2022gqt}, focusing on theoretical issues needed for future developments, and very extensively as well within the energy frontier working group EF05~\cite{Campbell:2022qmc}. More details regarding current status and necessary future developments can be found within those whitepapers.

\subsection{Theoretical developments in the SMEFT beyond dim-6}

The main goal of the precision program in particle physics is to uncover signatures of new physics to reveal a more fundamental theory 
beyond the SM. One way to guide experimental searches for new phenomena is to propose complete models of new physics with ultraviolet completions. Examples of this approach include supersymmetric versions of the SM with various mechanisms for soft SUSY breaking, or technicolor theories. Another approach, which is becoming increasingly used as LHC searches continue to return null results, is to write down an effective theory that incorporates a wide variety of possible UV models. One advantage of this method is that it is agnostic to high-energy details of UV completions and instead focuses on the dynamics at lower-energies that are relevant for current experimental searches. These EFTs may contain new light degrees of freedom that solve issues of the SM such as the need for a dark matter candidate, or they may contain only the observed SM states, implying a mass gap between the SM states and any new physics. If the discovered Higgs particle arises completely from an underlying SU(2) doublet, the resulting EFT is known as the Standard Model EFT (SMEFT). When the Higgs boson does not necessarily arise completely from an SU(2) doublet, the resulting theory is known as the Higgs EFT (HEFT) or the electroweak chiral Lagrangian. The HEFT is extensively reviewed in a number of papers~\cite{Feruglio:1992wf,Alonso:2012px,Brivio:2016fzo}. Since there 
is currently no evidence for a deviation of Higgs properties from those predicted for an SU(2) doublet, we focus here on the SMEFT.

The SMEFT Lagrangian is constructed containing only the SM degrees of freedom, and assuming that all operators satisfy the SM gauge symmetries. This leads to a result that differs from the SM Lagrangian by a series of higher-dimensional operators:
\begin{equation}
{\cal L}_{\text{SMEFT}} = {\cal L}_{\text{SM}}+\sum_{d=5}^{\infty} \sum_i \frac{1}{\Lambda^{d-4}} C_{d,i} {\cal O}_{d,i}.
\label{eq:smeftlag}
\end{equation}
Here, $\Lambda$ denotes a high energy scale at which the EFT description breaks down. $d$ denotes the dimension of the operator ${\cal O}_{d,i}$. The Wilson coefficients $C_{d,i}$ encode the dynamics of the UV completion. The index $i$ runs over all operators that appear at a given dimension. Predictions in this EFT require not only expansions in the usual SM couplings, but also in the operator dimensions that appear in Eq.~(\ref{eq:smeftlag}). The EFT contains two expansion parameters: the ratio $v/\Lambda$, where $v$ denotes the vacuum expectation value of the Higgs field and is representative of the SM particle masses, and $E/\Lambda$, where $E$ is the characteristic energy scale of the experimental processes under consideration. As experimental precision increases, and as $E$ increases, higher operator powers must be considered for reliable predictions. The leading dimension-5 correction in Eq.~(\ref{eq:smeftlag}) contains the Majorana neutrino mass term that violates lepton number. The UV scale associated 
with this operator is usually taken to be approximately $10^{13}$ GeV to reproduce the observed neutrino masses, and is therefore not relevant for current or planned future collider experiments. The same statement holds for all odd-dimensional operators. The complete, non-redundant basis of dimension-6 operators has been known for over a decade now~\cite{Buchmuller:1985jz,Arzt:1994gp,Grzadkowski:2010es}, and is extensively used as a framework for global fits of LHC and other experimental data. It is the study of the dimension-eight and higher terms that go beyond the first order in the SMEFT expansion, needed for precision studies, that is a natural topic for the TF06 working group.

The theoretical advances needed to understand dimension-8 and beyond terms in the SMEFT was summarized as part of the TF06 working group in the white paper~\cite{Alioli:2022fng}. Only a brief outline of the relevant work is presented here. The reader is referred to the white paper~\cite{Alioli:2022fng} for more details and complete lists of references.
\begin{itemize}

\item Predictions at dimension-8 and higher require an understanding of the relevant operator basis, a non-trivial task whose difficulty is 
highlighted by the fact that the independent operator basis for dimension-6 was not understood until decades after the first attempt at writing it down. The problem of counting operators at a given dimension was solved using Hilbert series techniques, and an understanding of the role played by the conformal group in incorporating constraints from integration-by-parts identities and field redefinitions~\cite{Lehman:2015via,Lehman:2015coa,Henning:2015daa,Henning:2015alf}. The explicit 
dimension-8 operator basis was recently derived~\cite{Murphy:2020rsh,Li:2020gnx}.

\item For $1 \to 2$ scattering processes, the structure of the SMEFT expansion is simple enough to allow for predictions to all orders in the $1/\Lambda$ expansion. This all-orders solution has a geometric interpretation, and is titled ``geoSMEFT~\cite{Helset:2020yio}." All-orders results for certain amplitudes can also be derived within the on-shell approach to SMEFT~\cite{Durieux:2019eor}. A main goal of this program is to learn from the available exact results how to better estimate truncation uncertainties in EFT analyses when higher-order corrections in $1/\Lambda$ are not derived~\cite{Martin:2021vwf}. 

\item The renormalization group equations governing the running of the dimension-6 SMEFT Wilson coefficients have been completely determined~\cite{Jenkins:2013zja,Jenkins:2013wua,Alonso:2013hga}. Accounting for these effects is important when combining results from experiments that span energy scales. The determination of these effects at dimension-8 is still in its infancy. A study of the effects from pairs of dimension-6 insertions in loops, which contribute at the same order as dimension-8 effects, was recently performed~\cite{Chala:2021pll}. The running effects at ${\cal O}(v^4/\Lambda^4)$ have a significant impact on positivity bounds on Wilson coefficients in the SMEFT~\cite{Chala:2021wpj}.

\item The phenomenological impact of ${\cal O}(v^4/\Lambda^4)$ effects can be significant, especially when the measurement is extremely precise or the experimental energy is large. Fits to the electroweak precision data can shift significantly upon inclusion of quadratic insertions of certain dimension-6 operators~\cite{Corbett:2021eux}. Inclusion of genuine dimension-8 effects can lead to novel angular dependence in Drell-Yan production~\cite{Alioli:2020kez}, as well as to shifts of the transverse momentum spectrum not possible at dimension-6~\cite{Boughezal:2022nof}, and can significantly change the interpretation of dimension-6 bounds obtained using LHC data~\cite{Boughezal:2021tih}. Novel low-energy phenomena such as toroidal quadrupole moments of the deuteron or positronium can be induced by dimension-8 operators~\cite{Mereghetti:2013bta}, and low-energy experiments can generally help disentangle dimension-6 from dimension-8 effects in fits~\cite{Boughezal:2021kla}. There have also been recent studies of the explicit form that dimension-8 corrections take in UV models~\cite{Dawson:2022cmu,Cohen:2020qvb}.

\end{itemize}

\section{Precision flavor physics}

\subsection{Introduction}

A key feature of flavor physics --- the study of interactions that distinguish between the three generations, i.e., break the global $[SU(3)]^5$ symmetry on the standard model --- is the plethora of observables that probe very high mass scales, well beyond the center-of-mass energy of the LHC or future colliders.  This is because quantum effects allow virtual particles to modify the
results of precision measurements.
Flavor physics can teach us about (multi-)TeV-scale new
physics, which cannot be learned from the direct production of new particles. 
The high mass-scale sensitivity arises because the SM flavor structure implies strong suppressions of flavor-changing neutral-current (FCNC)
processes (by the GIM mechanism, loop factors, and CKM elements). 
Even as the LHC continues to directly probe the TeV scale, ongoing and planned flavor physics experiments are sensitive to beyond standard model (BSM) flavor-changing interactions at much higher mass scales.  
These experiments provide essential constraints and complementary information on models proposed to explain any discoveries at the LHC or future colliders, and they have
the potential to reveal new physics that is inaccessible at the energy frontier.

Throughout the history of particle physics, studies of rare
processes, especially flavor-changing neutral currents (FCNCs) and $CP$ violation, have led to new and deeper understanding of Nature.  Weak interactions
foretold the electroweak scale.  Kaon decay experiments
were crucial for the development of the standard model: the discovery of $CP$
violation in $K_L^0 \to \pi^+ \pi^-$ decay ultimately pointed toward the
three-generation CKM model~\cite{Kobayashi:1973fv, Cabibbo:1963yz}; the absence of strangeness-changing
neutral current decays (i.e., the suppression of $K_L^0 \to \mu^+ \mu^-$ with
respect to $K^+ \to \mu^+ \nu$) led to the prediction of the fourth (charm)
quark~\cite{Glashow:1970gm}, and the measured value of the $K_L$\,--\,$K_S$ mass difference
made it possible to predict the charm quark mass~\cite{Gaillard:1974hs, Vainshtein:1973md}
before charm particles were detected.  More recently, the larger than
expected $B_H$\,--\,$B_L$ mass difference foretold the high mass of the top
quark.  Precision measurements of time-dependent $CP$ asymmetries in
$B$-meson decays in the BaBar and Belle experiments established the SM
as the leading source of $CP$ violation observed to date in flavor-changing
processes, leading to the 2008 Nobel Prize for Kobayashi and Maskawa.  
Nevertheless, corrections to the SM in FCNC processes at the tens of percent level are still allowed, and extensions of the SM are strongly constrained by flavor physics measurements and may have observable signals in the next generation of experiments.
(See Refs.~\cite{Ligeti:2015kwa, Grossman:2017thq, Gori:2019ybw, Grossman:2021xfq} for recent reviews.)

In the next one to two decades, the LHC experiments~\cite{Bediaga:2018lhg, Cerri:2018ypt}, Belle~II~\cite{Belle-II:2018jsg}, BES\,III~\cite{BESIII:2022mxl} and their planned and possible upgrades will increase $b$-, $c$-, and $\tau$-decay data sets by about two orders of magnitude.  
As in the past, theory will be crucial for the interpretation of the measurements and for maximizing their sensitivity to BSM physics, and experimental results will be essential inputs of theory considerations and triggers for new developments.

\subsection{Flavor probes of new physics}

\begin{figure}[tb]
\centerline{\includegraphics[width=.62\textwidth]{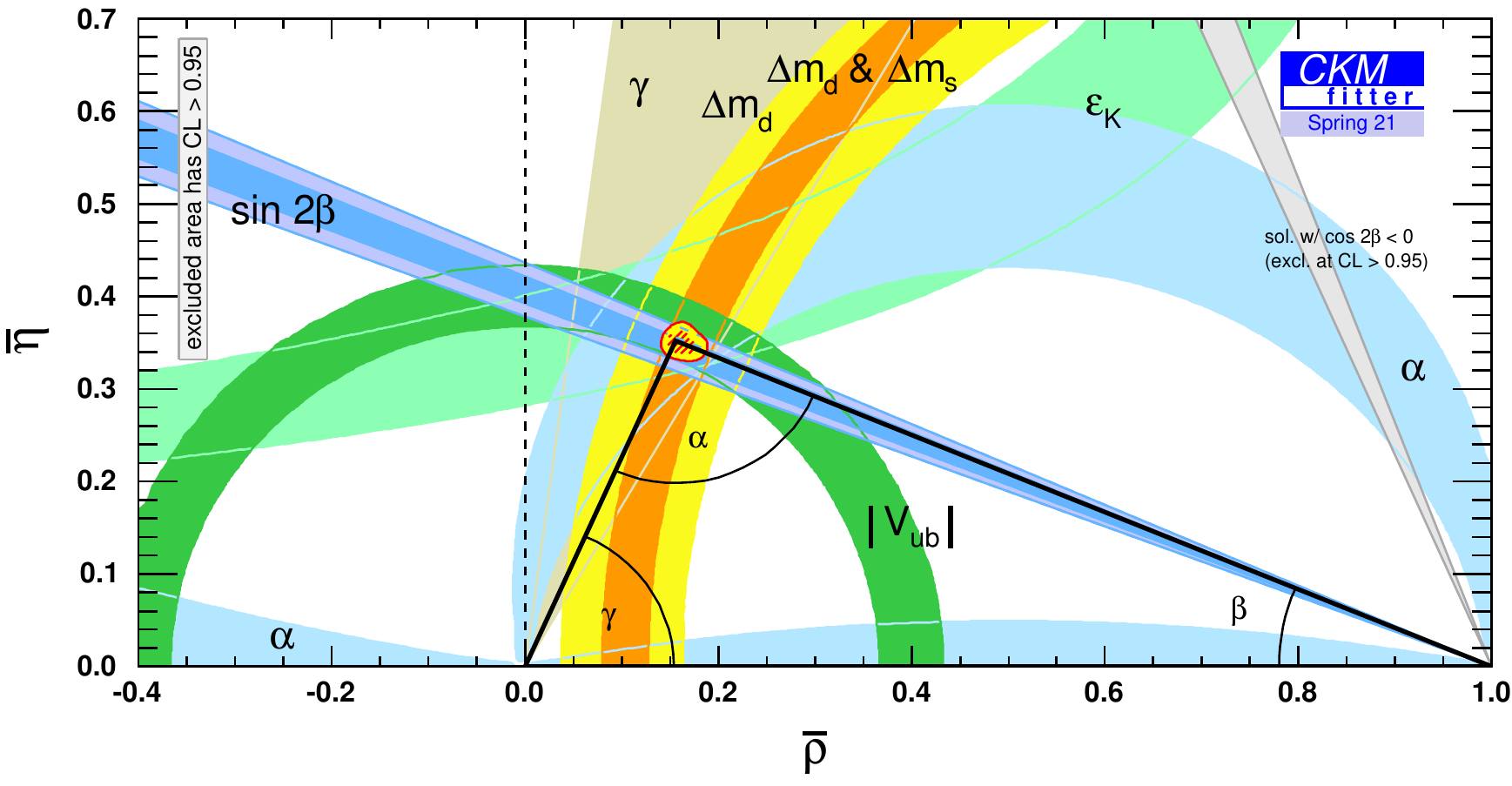} \hfil
\raisebox{5pt}{\includegraphics[width=.38\textwidth]{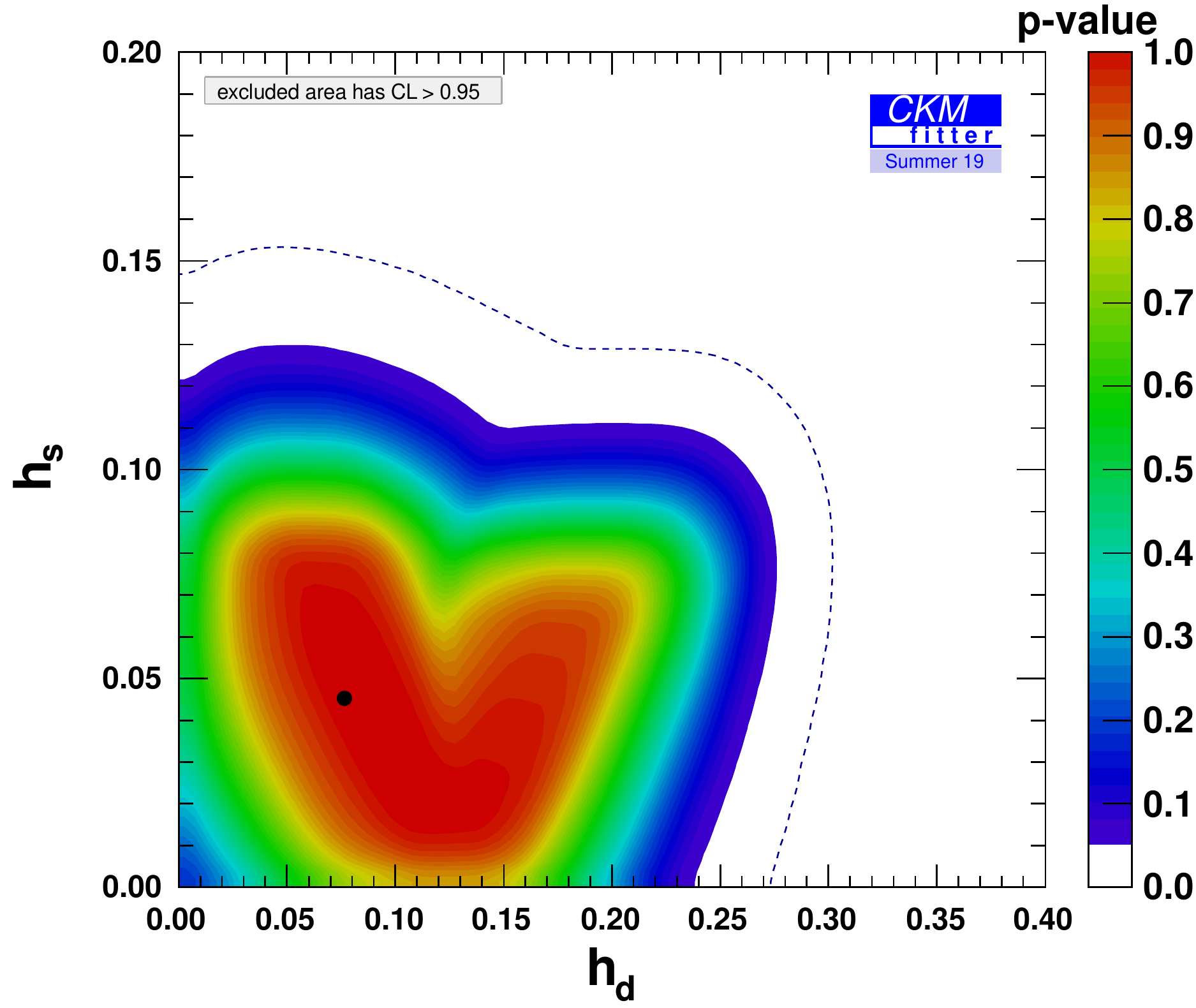}}}
\vspace*{-8pt}
\caption{Left: Constraints on the CKM parameters $\bar\rho$ and $\bar\eta$ in the SM (at 95\% CL)~\cite{Charles:2004jd}.  
Right: The allowed range of $h_d$ and $h_s$, the magnitudes of the BSM contributions to $B^0$ and $B_s^0$ mixing relative to the SM (see text)~\cite{Charles:2020dfl}. The black dot shows the best-fit point.}
\label{fig:ckmfit}
\end{figure}

The measurements of dozens of $CP$-violating and FCNC processes at $e^+e^-$ colliders and at the LHC are consistent with the SM predictions, with ever-increasing precision (see Fig.~\ref{fig:ckmfit}, left plot).  
(A few exceptions, where significant anomalies occur, are discussed below.)  
This strengthened the ``new physics flavor problem", which is the
tension between the hierarchy puzzle motivating BSM physics near the electroweak scale, and the high scale that is seemingly required to suppress BSM contributions to flavor-changing processes.  
The higher the scale of new physics, the more general its flavor structure may be, and the less clearly it can help with the hierarchy puzzle.
Improving constraints on the deviations from the SM predictions both in the properties of the Higgs particle and in the flavor sector, only makes the puzzle stronger;
TeV-scale NP entering even at the same loop order and with SM-like CKM couplings is being constrained.

As a simple example, relevant for a large class of models, assume that the dominant effect of BSM physics is to modify the mixing amplitudes of
neutral mesons, and leave tree-level decays unaffected.  This can be
parameterized by two real parameters for each neutral meson system.  The mixing of $B_q^0$ mesons (where $q=d,s$) are simplest to analyze, as they are dominated by short-distance physics.  Writing the mixing amplitude as $M_{12}^q = M_{12\, \rm (SM)}^q\, \big(1 + h_q\,
e^{2i\sigma_q}\big)$, the constraints on $h_d$ and $h_s$, the magnitudes of the BSM contributions relative to the SM, are shown in the
right plot in Fig.~\ref{fig:ckmfit}.  It shows that order $10-20\%$ corrections to $M_{12}$ are
still allowed (evidence for $h_q \neq 0$ would rule out
the SM).  Similar conclusions apply to other neutral meson
mixings, as well as many other $\Delta F=1$ FCNC
decays (e.g., $B\to X\gamma$, $B\to X\ell^+\ell^-$, $B_{d,s}\to \ell^+\ell^-$, $K\to \pi\nu\bar\nu$, etc.).

\begin{figure}[t]
\centerline{\includegraphics[width=.65\textwidth]{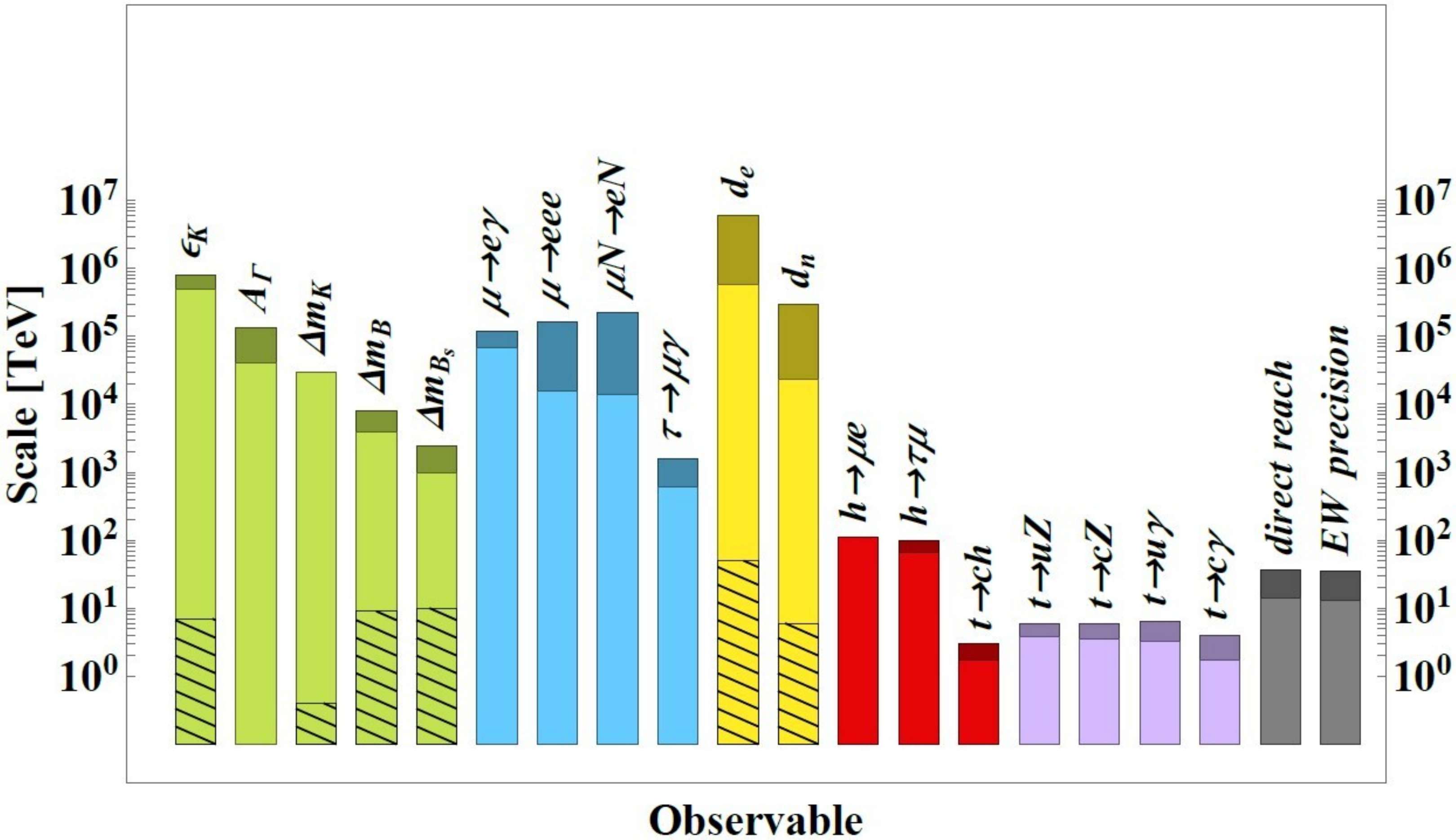}}
\vspace*{-5.75cm}
{\small \hspace*{4.5cm}\color{Green}{mesons}
\hspace*{.6cm}\color{blue}{leptons}
\hspace*{.3cm}\color{YellowOrange}{EDM}
\hspace*{.2cm}\color{red}{higgs} 
\hspace*{.5cm}\color{Fuchsia}{top}}
\vspace*{5.25cm}
\caption{Scales of dimension-6 operators probed by various observables assuming unit couplings.
The shaded regions show estimated future improvements, and the hatched regions show the scales probed in models with MFV flavor structure.
(From Ref.~\cite{EuropeanStrategyforParticlePhysicsPreparatoryGroup:2019qin}.)}
\label{fig:scales}
\end{figure}

The bounds from the consideration of a greater variety of $CP$-violating and flavor-changing observables are shown in Fig.~\ref{fig:scales}, encompassing the quark-, lepton-, and scalar sectors~\cite{EuropeanStrategyforParticlePhysicsPreparatoryGroup:2019qin}.  The scale of dimension-6 operators are shown, assuming ${\cal O}(1)$ coupling strength at present (light shading) and with anticipated mid-term improvements (dark shading; including HL-LHC, Belle~II, MEG~II, Mu3e, Mu2e, COMET, ACME, PIK, and SNS).  The greatest improvements in mass-scale sensitivity in the next $10-20$ years, by a whole order of magnitude, are expected in $\mu$ to $e$ conversion ($\mu N \to eN$) and electric dipole moment (EDM) experiments.
The bounds require that new physics is either at a very high scale or involves tiny couplings; the hatched bars show how they are weaker in models with minimal flavor violation, where the loop- and CKM-suppressions of the SM also apply.
%The new physics flavor puzzle is thus the question of why, and in what way, the flavor structure of the new physics is non-generic. 

Some of these bounds, especially from the kaon sector, have been known since the 1960s (most significantly $\Delta m_K$ and $\epsilon$, and later $\epsilon'$), thus flavor constraints have always been an input and rarely an output or prediction of BSM model building.
With the development of new classes of BSM models, numerous mechanisms were invented to suppress new contributions to flavor-changing processes, and thereby render TeV-scale BSM scenarios not yet ruled out.

Regarding the impact on model building, in supersymmetric (SUSY) extensions of the SM, box diagrams with winos and squarks contribute to meson mixing, and the structure of their couplings depends on SUSY breaking.  Some of the mechanisms proposed to suppress the SUSY contributions include, degeneracy~\cite{Dine:1995ag}, quark-squark alignment~\cite{Nir:1993mx}, heavy (3rd generation) squarks~\cite{Cohen:1996vb}, Dirac gauginos with an extra $R$ symmetry~\cite{Kribs:2007ac}, split SUSY~\cite{Arkani-Hamed:2005zuc}, and many more~\cite{Fox:2005yp}.
There are specific models to address the new physics flavor problem in
extra-dimensional models as well~\cite{Arkani-Hamed:1999ylh}.
It was the severe FCNC constraints in technicolor theories that 
inspired the ansatz that BSM sources of the breaking of the global $[U(3)]^5$ symmetry of the SM, without Yukawa couplings, may be proportional to the same Yukawas --- minimal flavor violation (MFV) --- which has since been widely applied to BSM model building~\cite{Chivukula:1987py, Hall:1990ac, DAmbrosio:2002vsn}.
Since the SM already breaks the $[U(3)]^5$ flavor symmetry, MFV gives a framework to
characterize ``minimal reasonable" deviations from the SM.  The corresponding constraints are indicated by the lower hatched parts of the sensitivity-bars in Fig~\ref{fig:scales}, and show that BSM scenarios with MFV flavor structures have no signals in many of these observables, and reduced sensitivity in the one to tens of TeV range in others.
Many other BSM scenarios are reviewed in the whitepaper~\cite{Altmannshofer:2022aml}.  Finding other paradigms that enable flavor-safe BSM model building would be influential.

\subsection{Flavor developments broadly impacting theory}

The richness of $B$ physics and the large $b$-quark mass ($m_b\gg\lqcd$) enables many complementary tests of the SM, and has been a driving force to develop new perturbative multi-loop and nonperturbative effective field theory techniques since the 1980s (see, e.g., Ref.~\cite{Buras:2011we}).  An iconic example is $B\to X_s\gamma$, which receives large and calculable QCD corrections~\cite{Bertolini:1986th, Grinstein:1987vj} and is also sensitive to BSM contributions~\cite{Hou:1987kf, Grinstein:1987pu}.  It has become one of the most complex SM calculations, with some parts of it involving the evaluation of 3-loop matching at the electroweak scale, 4-loop running, and 3-loop matrix elements~\cite{Misiak:2006zs, Misiak:2015xwa}.  In addition, only a restricted part of the photon energy phase space is accessible experimentally, and the well-measured part is subject to nonperturbative effects related to the $b$-quark distribution function in the $B$ meson~\cite{Neubert:1993um, Bigi:1993ex}.  This makes it essential to consistently 
combine fixed-order perturbative calculations, resummations of large logs (of various kinds) in endpoint regions, and nonperturbative ingredients~\cite{Ligeti:2008ac, Bernlochner:2020jlt}.  Addressing each of these issues has grown into significant areas of research.

The desire for (hadronic) model-independent understanding of semileptonic $B\to D^{(*)}\ell\bar\nu$ decays led to the development of the heavy quark effective theory (HQET)~\cite{Isgur:1989vq, Isgur:1990yhj, Georgi:1990um}.  
The techniques developed in that context were in turn instrumental in the development of many other effective field theories.  
For example, HQET had a significant impact on the development of NRGR~\cite{Goldberger:2004jt}, although instead of an HQET-like formulation with time-dependent velocity labels for the fields in a $g_{\mu\nu}$ background, it was simpler to formulate the EFT using worldlines~\cite{Walter} (as in an early formulation of HQET~\cite{Politzer:1988bs}).
Studying inclusive $B\to X\ell\bar\nu$ decays led to the development of the heavy quark expansion~\cite{Chay:1990da, Bigi:1992su, Bigi:1993fe, Manohar:1993qn}, which has served as a model for other operator product expansions.
The soft-collinear effective theory SCET~\cite{Bauer:2000ew, Bauer:2000yr} was developed initially motivated by summing Sudakov logs in $B\to X_s\gamma$ in an effective theory framework.  First applications included form factor relations in exclusive heavy to light decays in the phase space region $q^2 \ll m_B^2$~\cite{Charles:1998dr, Dugan:1990de}.  Systematically exploring the previously found vanishing of the forward-backward asymmetry in $B\to K^*\ell^+\ell^-$ at a particular value of $q^2$~\cite{Burdman:1998mk} led to an explosion in theoretical and experimental studies of angular distributions in $B\to K^*\ell^+\ell^-$ and related decays.
QCD factorization~\cite{Beneke:1999br, Beneke:2000ry}, and its systematic understanding using SCET, has been and will remain crucial to understand nonleptonic decays and many $CP$ violating observables.
SCET has also become part of standard theory tool-kit for higher order collider physics calculations.

Heavy quark physics, particularly the renormalon problem and the unsuitability of the $\overline{\rm MS}$ mass below the scale $\overline{m}_Q$, prompted the development of various short-distance quark mass definitions~\cite{Bigi:1994ga, Czarnecki:1997sz, Beneke:1998rk, Hoang:1998ng, Hoang:1998hm, Hoang:1999zc}, critical for precise calculations of inclusive $B$ decays and threshold problems.  They will also be crucial for the determination of the top quark mass with the smallest possible uncertainty, should an $e^+e^-$ collider run at the $t\bar t$ threshold in the future.

In the area of multi-loop calculations, flavor physics instigated developments of numerous technical aspects~\cite{kirill}.  These include:
(i)~The Laporta algorithm~\cite{Laporta:2000dsw}, a generic way to solve integration-by-parts (IBP), that
eventually became a workhorse behind many multi-loop computations,
was developed in the context of the analytic computation of the electron $g-2$ at order $\alpha^3$~\cite{Laporta:1996mq};
(ii)~Applications of IBPs beyond the traditional {\tt Mincer} code was to a large extent due to flavor problems (such as the $\overline{\rm MS}$ to on-shell
mass relations, zero recoil $b\to c\ell\bar\nu$ calculations, etc.);
(iii)~Loop calculations for processes with massive particles in initial and/or final state, including integrals and IBPs, were driven by flavor physics problems
($B\to X_s \gamma$, semileptonic decays, etc.);
(iv)~The expansions ``by regions" applied to Feynman diagrams as a
tool to obtain physical results (in HQET- and NRQCD-like expansions)
were also developed and driven initially by computations for flavor problems.

There are many other examples that EFT tools, initially developed for flavor physics, find uses in other areas.
Precision calculations of WIMP-nucleon scattering rates were made more systematic and simpler using EFT techniques~\cite{Hill:2013hoa, Hill:2014yka, Hill:2014yxa, Chen:2018uqz} compared to not using those~\cite{Hisano:2015rsa}.  
Methods borrowed from SCET were used to derive precise predictions for the photon spectrum resulting from wino dark matter annihilation~\cite{Baumgart:2017nsr, Baumgart:2018yed}, and resummations all the way up to the Planck scale~\cite{Bauer:2020jay} may have significant effects.
In the context of Higgs measurements, predictions for exclusive $H\to J/\psi\,\gamma$ and related decays use techniques developed in flavor physics, and may be a promising way to eventually probe the charm Yukawa coupling.

\OMIT{
In terms of fitting tools and event generators, it is worth noting that the two main CKM fitting groups, CKMfitter~\cite{Charles:2004jd,Hocker:2001xe} and UTfit~\cite{UTfit:2005ras,UTfit:2007eik}, are collaborations of theorists and experimentalists, as expertise in both areas are indispensable for the interpretations of measurements that contribute to constraining the CKM parameters or new physics couplings.
}

\OMIT{\bf (ZL: Mention hepfit, gambit, etc., somewhere?)}

\subsection{Semileptonic decays}

Semileptonic decays are important both for more precise determinations of CKM elements in the SM, and because of their sensitivity to BSM physics.  The uncertainty in the determination of $|V_{cb}|$ is one of the largest parametric uncertainties in the SM predictions for FCNC rates involving $V_{ts}$ and/or $V_{td}$, such as $K\to\pi\nu\bar\nu$, $\epsilon_K$, $\Delta m_B$, $B\to X\gamma$, $B_{d,s}\to \mu^+\mu^-$, $B\to X \ell^+\ell^-$.
The uncertainty of $|V_{ub}|$ is of central importance to BSM searches, too, since together with the angle $\gamma$ they form a tree-level ``reference'' determination of the unitarity triangle, with which results of other measurements impacted by loop processes can be compared.
For inclusive decays, the operator product expansion will likely remain the main tool, while
exclusive decays will likely rely on lattice QCD calculations, and extending those for heavy-to-light decays to the full ranges of $q^2$ would make big impacts.  
Since this Topical Group is about the role of precision theory, while the lepton flavor universality violating (LFUV) anomalies motivate both theory and experiment at present, we only describe the most-often discussed observables for context.  We focus on what will make their SM predictions the most precise, and refer to the Snowmass whitepaper on flavor model building~\cite{Altmannshofer:2022aml} for a review of models that can accommodate them.

%current anomalies.  The most often discussed leptoquark and $Z'$ models, and many other scenarios are reviewed in 

Most prominent among the anomalies are the hints for LFUV in two set of processes.  One is in neutral current $b\to s\ell^+\ell^-$ transitions, $R_{K^{(*)}} = {\cal B}{(B\to K^{(*)}\mu^+\mu^-)} / {\cal B}{(B\to K^{(*)}e^+e^-)}$, where the most recent LHCb measurement shows a $3.1\sigma$ deviation from the SM, 
$R_K(1.1 < q^2 < 6.0\, \GeV^2) = 0.846^{+0.042}_{-0.039}{} ^{+0.013}_{-0.012}$~\cite{LHCb:2021trn}.  The $R_{K^*}$ results with the full Run~1--2 data sets are not yet available, and neither is an average by experimentalists for the significance of the deviation from the SM.  Recent fits by theorists~\cite{Geng:2021nhg, Altmannshofer:2021qrr, Alguero:2021anc} quote the significance as above $4\sigma$.
The other presently most significant anomaly is that in charged current $b\to c\ell\bar\nu$ transitions, $R(D^{(*)}) = {\cal B}{(B\to D^{(*)}\tau\bar\nu)} /
{\cal B}{(B\to D^{(*)} l\bar\nu)}$, where $l=e,\mu$ (BaBar and Belle measurements use the $e,\mu$ average, whereas LHCb uses the $\mu$ mode).
In this case the significance for the deviation from the SM is quoted as
$3.1\sigma$~\cite{Amhis:2019ckw} to $3.6\sigma$~\cite{Bernlochner:2021vlv}, depending on the treatment of correlations.  In both of these cases, the present data hint at about $15-20\%$ corrections to the SM predictions.

While at the current central values, more precise measurements of $R_{K^{(*)}}$ could establish a breakdown of the SM without theory input, for the $R(D^{(*)})$ measurements theory is critical in assessing as to whether there is a deviation from the SM.
In many models inspired by these hints of LFUV, there
are correlated deviations from the SM predictions in transitions mediated by
operators with flavor structures $b\bar s \ell^+\ell^-$ and $b\bar s
\nu\bar\nu$, which makes, for example, searching for $B\to K^{(*)}\nu\bar\nu$ particularly interesting.  The BSM model building spurred by these anomalies has also led to many new ATLAS and CMS searches~\cite{Altmannshofer:2022aml}.

The role of lattice QCD in exclusive semileptonic $B$ decay calculations will be crucial, especially for the determination of CKM elements and refining predictions for $R(D^{(*)})$.  It will take some time before LQCD can make a big impact on $R_{K^*}$, as in heavy-to-light decays the small-$q^2$ region corresponds to large $K^{(*)}$ momenta in the $B$ restframe, which is hard to compute on the lattice.  Recently new ideas emerged for lattice calculations of inclusive $B$ decays~\cite{Hashimoto:2017wqo}.
These and other expected developments are summarized in the whitepaper~\cite{Boyle:2022uba} and the TF05 report.

A common challenge in both $b\to q\ell\bar\nu$ and $b\to q\ell\bar\ell$ mediated decays is that going forward, the role of 
electromagnetic corrections needs to be better understood, and the related uncertainties assessed.  Recent work has focused on $b\to q\ell\bar\ell$~\cite{Bordone:2016gaq} inspired by $R_{K^*}$, and there are also ongoing efforts to understand the roles of photon radiation~\cite{Robinson:2021cws, Isidori:2022bzw, Isidori:2020acz} and how accurate {\tt PHOTOS}~\cite{Barberio:1990ms, Golonka:2005pn, Davidson:2010ew} is in describing it. 
More broadly, in both semileptonic and nonleptonic decays, a better understanding of electromagnetic radiative corrections, as well as all sources of isospin violation, are expected to become increasingly important.
Initial steps toward systematically including isospin violation and electromagnetic corrections in lattice QCD calculations have also started and will be increasingly relevant in the future~\cite{Sachrajda}.

\paragraph{$b\to q\ell\bar\nu$ mediated decays}

Semileptonic $B$ decays have continued to receive significant attention due to the importance of the determinations of $|V_{cb}|$ and $|V_{ub}|$, and also because there have been persistent tensions between their measurements from inclusive and exclusive semileptonic decays.  Experimental improvements will come from cleaner measurements with large Belle~II data sets, and measurements of the ratio $|V_{cb} / V_{ub}|$ at LHCb from exclusive decays.

There is a lot of ongoing work to improve our theoretical knowledge.  
Recently the order $\alpha_s^3$ corrections to the inclusive semileptonic $B\to
X_c\ell\bar\nu$ decay rate have been computed~\cite{Fael:2020tow}, as well as the 
3-loop corrections to the relations between the kinetic heavy quark mass and other quark mass definitions~\cite{Fael:2020njb}.  The impact of these on 
$|V_{cb}|$ have started to be assessed~\cite{Bordone:2021oof}.  
Already in the early 2000s, the $\lqcd^2/m_b^2$ corrections to measurable differential spectra were known, and (its moments) fitted to experimental measurements to determine simultaneously $|V_{cb}|$ and the hadronic matrix elements.  The nonperturbative corrections to the inclusive $B\to
X_c\ell\bar\nu$ rate have been worked out to $1/m_b^5$~\cite{Mannel:2010wj}, and recently some differential distributions have also been studied at order $1/m_b^3$ and $\alpha_s$~\cite{Mannel:2021zzr}.
In the future, such perturbative and nonperturbative corrections to the OPE may be calculated for differential rates as well, which, when combined with future data, could constrain higher-order terms in the heavy quark expansion more than ever done before, and simultaneously determine $|V_{cb}|$.
In the meanwhile, other strategies also emerge, such as the observation that due to reparametrization invariance
the $q^2$ spectrum has simpler structure at higher orders in the $\lqcd/m_{c,b}$ expansion than other distributions\cite{Fael:2018vsp}, allowing for complementary determinations of $|V_{cb}|$~\cite{Bernlochner:2022ucr}.

Recently, with the first publications of ``unfolded'' measurements of differential $B\to D^{(*)}\ell\bar\nu$ decay distributions from Belle, it became possible for theorists to perform fits to the data using different assumptions.  This has allowed testing different implementations of constraints on the shapes of form factors from analyticity and unitarity, and assessing the roles of model-dependent QCD sum rule inputs on prior measurements of $|V_{cb}|$ from these exclusive decays.  It also allows theorists to find possible new manifestations of LFUV, such as the recent $4\sigma$ claim~\cite{Bobeth:2021lya} of an $e$ vs.\ $\mu$ LFUV in an angular distribution.

Regarding the $R(D^{(*)})$ measurements and their tension with the SM, a lot of effort has gone into refining the SM predictions.  
This is a prime example of observables for which no useful statement could be made about the (in)consistency of the data with the SM without theory input.
If new physics contributes to these observables, then it also affects the measurements made using SM expectations for decay distributions and experimental efficiencies.  Thus, fitting the measurements based on the SM to new physics models containing different operators is of limited reliability.  The {\tt Hammer}~\cite{Bernlochner:2020tfi} tool allows reweighting event samples to arbitrary NP scenarios or to any hadronic matrix elements.

Concerning connections with SMEFT and HEFT, it was pointed out recently that the $R(D^{(*)})$ anomaly may be possible to accommodate via the operator
$(\bar{c} \gamma_\mu P_R b)\,(\bar\tau \gamma^\mu P_L \nu)$ in HEFT~\cite{Burgess:2021ylu}, while this was known to be difficult in SMEFT.  
The EFT formulation of low-energy observables below the electroweak scale (LEFT)~\cite{Jenkins:2017jig, Jenkins:2017dyc} would allow systematic and model independent interpretations of combinations of flavor physics and high-$p_T$ anomalies.

\paragraph{$b\to q\ell^+\ell^-$ mediated decays}

Among FCNC $B$ decays with the largest rates, besides $B\to X_s\gamma$, the $b\to q\ell^+\ell^-$ mediated decays have also been known for decades to provide complementary sensitivity to new physics, and a richer set of observables in the simplest exclusive decay channels, $B\to K^{(*)} \ell^+\ell^-$.  While the inclusive rate is calculable in an OPE, the comparison of theory and experiment for exclusive decays must rely on form factor calculations.  An exception is LFUV ratios of decay rates, which were hardly discussed in the theory literature before the first LHCb measurement of $R_K$~\cite{LHCb:2014vgu}.

While the hints of LFUV in $R_{K^{(*)}}$ are theoretically very clean to interpret as a breakdown of the SM at the current level of sensitivity, there are a number of other observables in these and related decays, where significant deviations from the SM predictions have been claimed.  
A central question for heavy-to-light decays is our ability to predict the form factors parametrizing each decay.  For many channels,
lattice QCD calculations exist at high $q^2$ and light-cone QCD sum rules at small $q^2$.  How reliable these calculations are, and the interpolations to connect them, are somewhat open questions.  For example, LHCb measured
${\cal B}(B_s\to \phi\mu^+\mu^-)(1.1 < q^2 < 6.0\, \GeV^2)$
at a rate that is $3.6\sigma$ below a SM prediction~\cite{LHCb:2021zwz}, based on a combination
of light-cone sum rule~\cite{Bharucha:2015bzk} and lattice QCD~\cite{Horgan:2013pva} inputs.  With more data and cross-checks, we will learn if this is a breakdown of these calculations or of the SM.

Concerning more differential observables in $B\to K^* \ell^+\ell^-$, there have been attempts since the 1990s to find BSM-sensitive but hadronic physics insensitive observables, such as the $q^2$ value at which the forward-backward asymmetry vanishes.  This became more systematic when SCET developments~\cite{Bauer:2002aj, Beneke:2003pa} put the form factor relations, first found in LEET, on more solid footing.  The form factors can be written as a nonperturbative part, which obeys symmetry relations, and at leading order depend on only two functions of $q^2$, and a hard scattering part, which breaks the symmetry relations, but is computable in an expansion in $\alpha_s$.
When studying observables that describe the full angular distribution in $B\to K^*(K\pi) \ell^+\ell^-$~\cite{Matias:2012xw}, of which $P_5'$ became best known due to tensions between SM predictions and data~\cite{LHCb:2020gog}, detailed theory input is needed to assess whether the data are (in)consistent with the SM.
Two main sources of theoretical uncertainties are subject to ongoing research and healthy debates.  One relates to estimating subleading corrections to the heavy-to-light form factor relations, not fully calculable from first principles at present.  The other relates to the role of $c\bar c$ loop contributions to the $\ell^+\ell^-$ invariant mass spectrum~\cite{Beneke:2009az}.  Most important is the region $q^2 < 6\,\GeV^2$; while seemingly far below $m_{J/\psi}^2$, given $\Gamma_{J/\psi} \simeq 93\,{\rm keV}$, a Breit-Wigner description is not sufficient.
In the $q^2 > m_{\psi'}^2$ region, factorization is known to be a poor approximation~\cite{Lyon:2014hpa}.
Future progress is much desired, and will likely be relevant for a better understanding of nonleptonic decays, too.

Recent fits to (certain subsets of) this data yield combined deviations from the SM that are quoted at least at the $4\sigma$ level~\cite{Geng:2021nhg, Altmannshofer:2021qrr, Alguero:2021anc}.
Prompted by the many measurements relevant for constraining BSM scenarios, several fitting codes have been developed to aid connecting the experimental measurements with BSM models, such as flavio~\cite{Straub:2018kue}, EOS~\cite{vanDyk:2021sup}, and SuperIso~\cite{Arbey:2018msw}.

\subsection{Nonleptonic decays and \texorpdfstring{$CP$}{CP} violation}

To fully utilize the next generation of measurements, a better theoretical understanding of nonleptonic decays is much desired.  
To date, most $CP$ violation measurements have been performed for such decays, as they not only allow measurements of the CKM unitarity triangle angles, but also provide numerous probes of $CP$ violating BSM interactions.

Consider the well-known and important example, the determination of $\sin(2\beta)$ from the time-dependent $CP$ asymmetry in $B\to J/\psi K_S$ and related modes, which was the ``gold-plated" measurement for BaBar and Belle.  The theoretical uncertainty was known to be at the percent level, and was negligible for the past $B$ factories.  
The reason is that amplitudes with one weak phase dominate the decay, with deviations suppressed by $|V_{ub}V_{us}/(V_{cb}V_{cs})| \approx 0.02$ times a ratio of hadronic matrix elements which is expected to be well below unity.  (This ``$V_{ub}$ contamination" is often, and a bit confusingly, called ``penguin pollution".)  
The current world average is $\sin(2\beta) = 0.699 \pm 0.017$~\cite{ParticleDataGroup:2020ssz} and since the measurements will improve by a lot, constraining this uncertainty will become important.  The same question arises for the analogous mode for $B_s$ decay, the time-dependent $CP$ asymmetry in $B_s \to J/\psi \phi$, which is sensitive to BSM contributions to $B_s$ mixing.  

The contributions of these CKM-suppressed $b\to u\bar u s$ operators to the respective amplitudes can either be attempted to be calculated, or related to other observables by flavor symmetry relations.  
There is no consensus yet regarding robust estimations of the uncertainties of QCD factorization based calculations of these quantities~\cite{Frings:2015eva}.  
Factorization receives ${\cal O}(30\%)$ corrections even in some channels like $B\to D^{(*)}\pi$, where it is actually proven in the heavy quark limit at leading order in $\lqcd/m_{c,b}$.  
Some approaches combine $SU(3)$ flavor symmetry and diagrammatic assumptions.
In the absence of a rigorous way to estimate the uncertainties, it is unclear if such approaches can establish whether possible future tensions are due to BSM contributions or underestimated uncertainties.  
It is possible using only $SU(3)$ flavor symmetry to relate the $V_{ub}$ contamination to other observable decay rates~\cite{Ligeti:2015yma}, including the isospin violation in $B\to J/\psi K$ and $J/\psi\pi$, and a challenge is to disentangle that from the rate difference between $\Upsilon(4S)\to B^0\Bbar^0$ and $B^+B^-$~\cite{Jung:2015yma}.

Another important set of processes are the so-called ``penguin modes", such as $B\to \phi K_S$, $\eta^{(\prime)}K_S$, $\omega K_S$, $\pi^0 K_S$, etc.\ (and the analogous $B_s\to \phi \phi$). 
The interest in these modes is due to the fact that the amplitudes are dominated by one-loop contributions in the SM, with weak phases close to the $b\to c\bar c s$ operator, and BSM contributions could alter the SM amplitudes and affect the $CP$ asymmetries (which are expected to be close to $\sin2\beta$ in the SM).
Past works attempted both explicit calculations~\cite{Beneke:2005pu, Williamson:2006hb, Zupan:2007ca} or flavor symmetry based relations~\cite{Grossman:2003qp}, and it is unclear what theory approaches can match the uncertainties expected with the full LHCb and Bellee~II data sets.  Developing new model-independent methods to go beyond the current state of the art will be important for the interpretation of future data.

Charmless two-body $B$ decays have received ample attention, and will continue to do so.  It is known since the CLEO observation of $B\to K\pi$~\cite{CLEO:1997ree}, before $B\to \pi\pi$, that ``penguin" amplitudes are not nearly as suppressed relative to ``tree" amplitudes as previously expected.  Developments in SCET and QCD factorization have shed some light on this, while many open questions remain, not the least is whether there are any hints of BSM contributions.  To what extent charm loop contributions to such decays are reliably tractable using perturbative methods remains unresolved~\cite{Bauer:2004tj, Bauer:2005kd, Beneke:2004bn}.
The difference of direct $CP$ asymmetries, $A_{K^-\pi^0} - A_{K^-\pi^+} = 0.086 \pm 0.013$~\cite{ParticleDataGroup:2020ssz} has been puzzling since the early 2000s using factorization-based approaches (but not based on $SU(3)$ flavor symmetry alone).  
The reasons is that the dominant tree and penguin contributions coincide in these two modes, and the color-allowed and color-suppressed tree contributions to $ K^-\pi^0$ have a power-suppressed relative strong phase in SCET, hence one is lead to require annihilation and electroweak penguin contributions to play a substantial role (which are often assumed to be negligible).

\subsection{Charm and Kaon decays}

$CP$ violation in both charm mixing and decay are probes of QCD dynamics and BSM physics.  While mixing and FCNCs in $K$, $B$, and $B_s$ mesons are generated by loop diagrams with intermediate up-type quarks in the SM, in the $D$ sector intermediate down-type quarks are at play.  (Or, in SUSY, down-type squarks for $K$, $B$, and $B_s$, and up-type squarks for $D$ meson FCNCs.)

After an earlier hint of a much larger effect, direct $CP$ violation in $D$ decay was established in 2019 by LHCb, 
$A_{CP}(K^-K^+) - A_{CP}(\pi^-\pi^+) = -(1.54\pm0.29)\times 10^{-3}$~\cite{Aaij:2019kcg}.
This result can be accommodated in the SM, but it requires hadronic enhancements compared to the available (model dependent) calculations.  
$CP$ violation in $D$ mixing is thought to be a theoretically cleaner probe of the SM.  While the experimental bounds improved recently~\cite{LHCb:2021dcr}, there is probably still significant room for discovery~\cite{Kagan:2020vri}.  Future theory research needs to establish what the maximal size of $CP$ violation in $D$ mixing is, that could not be attributed to the SM and would signal new physics.
While we cannot at present make precision calculations because of 
hadronic uncertainties, one can explore consequences of flavor $SU(3)$ symmetry. Different observables arise at different orders of $SU(3)$ breaking, thus, by
studying relations that hold to higher order in $SU(3)$ breaking, we can test the picture of flavor symmetry breaking and understand a pattern that can be used to probe the SM.  
The situation can dramatically change if there is a theoretical breakthrough that enables the calculation of some of the hadronic inputs needed for these observables. The prime candidate is lattice QCD, which may be able to address nonleptonic $D$ decays well before addressing nonleptonic $B$ decays

Another active and interesting area is the study of FCNC $D$ decays~\cite{Bause:2020xzj}.  The challenge is to limit the role of long-distance contributions, and there are several promising directions.  For example, the lack of photon coupling to $\nu\bar\nu$ implies that $c\to u \nu\bar\nu$ mediated decays may be sensitive probes of BSM~\cite{Bause:2020xzj}.  In the next decade, LHCb, Belle~II and BES\,III are 
expected to measure many rare decays and $CP$ asymmetries with good precision (and measure them individually, without having to form differences, as in the LHCb measurement mentioned above).

The rare decays $K^\pm\to\pi^\pm\nu\bar\nu$ and especially $K_L
\to\pi^0\nu\bar\nu$ are sensitive to some of the highest scales in flavor
physics (in the $\Delta F=1$ sectors, i.e., outside of neutral meson mixings),
as the SM contributions are severely suppressed both by the GIM mechanism and
small CKM angles.  The $K^\pm\to\pi^\pm\nu\bar\nu$ rate is only known at the $\sim 40\%$ level to agree with the SM~\cite{NA62:2021zjw}, and $K_L
\to\pi^0\nu\bar\nu$ may still deviate from the SM prediction substantially.

Recently it was proposed that $K\to\mu^+\mu^-$ can also be a clean probe of short-distance
physics~\cite{DAmbrosio:2017klp, Dery:2021mct}.  While the total rate has large
long-distance contributions, the $CP$ violating part has a theoretically clean
interpretation in terms of short distance physics.  The new idea is that by doing time-dependent interference measurements, ${\cal
B}(K_S \to \mu^+\mu^-)_{\ell=0}$ can be determined, which has clean
interpretation in terms of short distance physics.  It allows testing the SM and
is sensitive to BSM physics.
There can be BSM contributions to $K\to\mu^+\mu^-$ without affecting the
$K\to\pi\nu\bar\nu$ decays to a detectable extent~\cite{Dery:2021vql}.

Possible future avenues for the kaon program are outlined in a contributed paper, Ref.~\cite{Goudzovski:2022vbt}.

\section{Outlook}

Precision measurements can unlock energy scales far above the direct reach of collider experiments. For example, $CP$-violating observables and FCNC processes allow energy scales in excess of tens of TeV to be probed. Along with this opportunity comes a challenge: precision calculations in the SM are needed to tap into the potential of these data sets. The power and lure of such data have proven over and over to stimulate both theorists and experimenters to invent creative groundbreaking methods, both in high-$p_T$ and in flavor physics, which enable new calculations and measurements at previously unexpected precision and sensitivity.

The future of HEP will benefit from large, exquisite data from the high-luminosity LHC and Belle II, as well as future colliders with even higher luminosity. As the opportunity becomes greater so does the challenge. Continued advances in high-precision theory are needed to maximize the discovery potential of these future experiments, including ever-higher order perturbative calculations, advanced Monte Carlo event generators and parton showers for a closer description of experimental data, and parton distribution functions extracted to a matching precision. The past and expected forthcoming impact of precision calculations on the
interpretation of experimental data demands this research to be well supported.

\subsection*{Acknowledgement}

We thank all the contributors to the TF06 working group for their input and white paper submissions.
We thank Walter Goldberger, Kirill Melnikov, Bernhard Mistlberger, Robert Szafron, and Jure Zupan for helpful discussions.
R.~B.\ is supported by the U.S.~Department of Energy (DOE) contract DE-AC02-06CH11357.
Z.~L.~was supported in part by the U.S.~Department of Energy under contract DE-AC02-05CH11231.
%
%%%% New %%%%
W.~A.\  is supported by the U.S.~Department of Energy grant number DE-SC0010107. 
S.~D.~B.\ is supported by the projects P18-FR-4314, and A-FQM-467-UGR18 (FEDER).
F.~C.\ was partially supported by the ERC Starting Grant 804394 {\sc hipQCD} and by the UK Science and Technology Facilities Council (STFC) under 
grant ST/T000864/1.
M.~C.\ is supported by the Ram\'on y Cajal programme as well as by the SRA grant PID2019-106087GB-C21/C22 (10.13039/501100011033) and the Junta de Andaluc\'ia grants FQM 101, A-FQM-211-UGR18 and P18-FR-4314 (FEDER).
A.~D.~C.\ is supported by MCIN/AEI under the grant PID2019-106087GB-C21 (10.13039/501100011033) and by ESF Investing in your future, as well as by Junta de Andaluc\'ia under grant FQM 101.
W.~C.\ is supported by the Natural Science Foundation of China (NSFC) under the contract No. 11975200. 
B.~H.\ is supported by the Swiss National Science Foundation, under grants no.~PP00P2-170578 and no.\ 200020-188671, and through the National Center of Competence in Research SwissMAP.
S.~J.\ is supported by the UK Science and Technology Facilities Council under grant ST/T001011/1.
T.~K.\ is supported by the JSPS Grant-in-Aid for Early-Career Scientists (Grant No.\ 19K14706).
H.-L.~L.\ is supported by F.R.S.-FNRS through the IISN convention "Theory of Fundamental Interactions” (N: 4.4517.08).
X.~L.\ is supported by the National Natural Science Foundation of China under Grant No. 12175016 and by the National Special Support Program for High-level Talents. 
A.~M.\ is supported by NSF fund no.\ PHY-2112540.
M.~R.~M.\ is supported by the UK Science and Technology Facilities Council (grant numbers ST/P001246/1). The work of M.~R.~M. has also received funding from the European Union’s Horizon 2020 research and innovation program as part of the Marie Sklodowska-Curie Innovative Training Network MCnetITN3 (grant agreement no. 722104).
T.~M.\ is supported by the World Premier International Research Center Initiative (WPI) MEXT, Japan, and by JSPS KAKENHI grants JP19H05810, JP20H01896, and JP20H00153. 
E.~M.\ is supported  by the U.S.~Department of Energy through the Office of Nuclear Physics  and  the  LDRD program at Los Alamos National Laboratory.
B.~M.\ is supported by the U.S.~Department of Energy, Contract DE-AC02-76SF00515.
F.~P.\ is supported by the U.S.~Department of Energy (DOE) contract DE-AC02-06CH11357 and DE-FG02-91ER40684.
D.~S.\ is supported in part by the U.S.~Department of Energy under grant DE-SC0011640.
G.~S.\ is supported by the National Science Foundation, awards 1915093 and 2210533.
R.~S.\ is supported by the U.S.~Department of Energy under Grant Contract DE-SC0012704.
L.~V.\ is supported by Fellini, Fellowship for Innovation at INFN, funded by the European Union’s Horizon 2020 research programme under the Marie Sk\l{}odowska-Curie Cofund Action, grant agreement no.~754496.
G.~V.\ is supported by the U.S.~Department of Energy, Contract DE-AC02-76SF00515.
J.-H.Y.\ is supported by the National Science Foundation of China under Grants No. 12022514, No.\ 11875003 and No.\ 12047503, and National Key Research and Development Program of China Grant No.\ 2020YFC2201501, No.\ 2021YFA0718304, and CAS Project for Young Scientists in Basic Research YSBR-006, the Key Research Program of the CAS Grant No.\ XDPB15. 
J.~Z.\ acknowledges support in part by the DOE grant DE-SC0011784 and NSF OAC-2103889.

% lists all entries in bib file(s)
%\nocite*
\bibliography{TF06a, TF06b}

\providecommand{\href}[2]{#2}\begingroup\raggedright\begin{thebibliography}{100}

\bibitem{Amoroso:2022eow}
S.~Amoroso {\em et~al.}, ``{Snowmass 2021 whitepaper: Proton structure at the
  precision frontier},'' \href{http://arxiv.org/abs/2203.13923}{{\ttfamily
  arXiv:2203.13923 [hep-ph]}}.

\bibitem{Anastasiou:2016cez}
C.~Anastasiou, C.~Duhr, F.~Dulat, E.~Furlan, T.~Gehrmann, F.~Herzog,
  A.~Lazopoulos, and B.~Mistlberger, ``{High precision determination of the
  gluon fusion Higgs boson cross-section at the LHC},''
  \href{http://dx.doi.org/10.1007/JHEP05(2016)058}{{\em JHEP} {\bfseries 05}
  (2016) 058}, \href{http://arxiv.org/abs/1602.00695}{{\ttfamily
  arXiv:1602.00695 [hep-ph]}}.

\bibitem{Dreyer:2016oyx}
F.~A. Dreyer and A.~Karlberg, ``{Vector-Boson Fusion Higgs Production at Three
  Loops in QCD},'' \href{http://dx.doi.org/10.1103/PhysRevLett.117.072001}{{\em
  Phys. Rev. Lett.} {\bfseries 117} no.~7, (2016) 072001},
  \href{http://arxiv.org/abs/1606.00840}{{\ttfamily arXiv:1606.00840
  [hep-ph]}}.

\bibitem{Duhr:2019kwi}
C.~Duhr, F.~Dulat, and B.~Mistlberger, ``{Higgs Boson Production in
  Bottom-Quark Fusion to Third Order in the Strong Coupling},''
  \href{http://dx.doi.org/10.1103/PhysRevLett.125.051804}{{\em Phys. Rev.
  Lett.} {\bfseries 125} no.~5, (2020) 051804},
  \href{http://arxiv.org/abs/1904.09990}{{\ttfamily arXiv:1904.09990
  [hep-ph]}}.

\bibitem{Chen:2019lzz}
L.-B. Chen, H.~T. Li, H.-S. Shao, and J.~Wang, ``{Higgs boson pair production
  via gluon fusion at N$^3$LO in QCD},''
  \href{http://dx.doi.org/10.1016/j.physletb.2020.135292}{{\em Phys. Lett. B}
  {\bfseries 803} (2020) 135292},
  \href{http://arxiv.org/abs/1909.06808}{{\ttfamily arXiv:1909.06808
  [hep-ph]}}.

\bibitem{Duhr:2021vwj}
C.~Duhr and B.~Mistlberger, ``{Lepton-pair production at hadron colliders at
  N$^{3}$LO in QCD},'' \href{http://dx.doi.org/10.1007/JHEP03(2022)116}{{\em
  JHEP} {\bfseries 03} (2022) 116},
  \href{http://arxiv.org/abs/2111.10379}{{\ttfamily arXiv:2111.10379
  [hep-ph]}}.

\bibitem{Duhr:2020seh}
C.~Duhr, F.~Dulat, and B.~Mistlberger, ``{Drell-Yan Cross Section to Third
  Order in the Strong Coupling Constant},''
  \href{http://dx.doi.org/10.1103/PhysRevLett.125.172001}{{\em Phys. Rev.
  Lett.} {\bfseries 125} no.~17, (2020) 172001},
  \href{http://arxiv.org/abs/2001.07717}{{\ttfamily arXiv:2001.07717
  [hep-ph]}}.

\bibitem{Duhr:2020sdp}
C.~Duhr, F.~Dulat, and B.~Mistlberger, ``{Charged current Drell-Yan production
  at N$^{3}$LO},'' \href{http://dx.doi.org/10.1007/JHEP11(2020)143}{{\em JHEP}
  {\bfseries 11} (2020) 143}, \href{http://arxiv.org/abs/2007.13313}{{\ttfamily
  arXiv:2007.13313 [hep-ph]}}.

\bibitem{Caola:2022ayt}
F.~Caola, W.~Chen, C.~Duhr, X.~Liu, B.~Mistlberger, F.~Petriello, G.~Vita, and
  S.~Weinzierl, ``{The Path forward to N$^3$LO},'' in {\em {2022 Snowmass
  Summer Study}}.
\newblock 3, 2022.
\newblock \href{http://arxiv.org/abs/2203.06730}{{\ttfamily arXiv:2203.06730
  [hep-ph]}}.

\bibitem{Mistlberger:2018etf}
B.~Mistlberger, ``{Higgs boson production at hadron colliders at N$^{3}$LO in
  QCD},'' \href{http://dx.doi.org/10.1007/JHEP05(2018)028}{{\em JHEP}
  {\bfseries 05} (2018) 028}, \href{http://arxiv.org/abs/1802.00833}{{\ttfamily
  arXiv:1802.00833 [hep-ph]}}.

\bibitem{Duhr:2019rrs}
C.~Duhr and L.~Tancredi, ``{Algorithms and tools for iterated Eisenstein
  integrals},'' \href{http://dx.doi.org/10.1007/JHEP02(2020)105}{{\em JHEP}
  {\bfseries 02} (2020) 105}, \href{http://arxiv.org/abs/1912.00077}{{\ttfamily
  arXiv:1912.00077 [hep-th]}}.

\bibitem{Weinzierl:2020fyx}
S.~Weinzierl, ``{Modular transformations of elliptic Feynman integrals},''
  \href{http://dx.doi.org/10.1016/j.nuclphysb.2021.115309}{{\em Nucl. Phys. B}
  {\bfseries 964} (2021) 115309},
  \href{http://arxiv.org/abs/2011.07311}{{\ttfamily arXiv:2011.07311
  [hep-th]}}.

\bibitem{Walden:2020odh}
M.~Walden and S.~Weinzierl, ``{Numerical evaluation of iterated integrals
  related to elliptic Feynman integrals},''
  \href{http://dx.doi.org/10.1016/j.cpc.2021.108020}{{\em Comput. Phys.
  Commun.} {\bfseries 265} (2021) 108020},
  \href{http://arxiv.org/abs/2010.05271}{{\ttfamily arXiv:2010.05271
  [hep-ph]}}.

\bibitem{Bourjaily:2022bwx}
J.~L. Bourjaily {\em et~al.}, ``{Functions Beyond Multiple Polylogarithms for
  Precision Collider Physics},'' in {\em {2022 Snowmass Summer Study}}.
\newblock 3, 2022.
\newblock \href{http://arxiv.org/abs/2203.07088}{{\ttfamily arXiv:2203.07088
  [hep-ph]}}.

\bibitem{Gehrmann-DeRidder:2005btv}
A.~Gehrmann-De~Ridder, T.~Gehrmann, and E.~W.~N. Glover, ``{Antenna subtraction
  at NNLO},'' \href{http://dx.doi.org/10.1088/1126-6708/2005/09/056}{{\em JHEP}
  {\bfseries 09} (2005) 056},
  \href{http://arxiv.org/abs/hep-ph/0505111}{{\ttfamily arXiv:hep-ph/0505111}}.

\bibitem{Currie:2013dwa}
J.~Currie, A.~Gehrmann-De~Ridder, E.~W.~N. Glover, and J.~Pires, ``{NNLO QCD
  corrections to jet production at hadron colliders from gluon scattering},''
  \href{http://dx.doi.org/10.1007/JHEP01(2014)110}{{\em JHEP} {\bfseries 01}
  (2014) 110}, \href{http://arxiv.org/abs/1310.3993}{{\ttfamily arXiv:1310.3993
  [hep-ph]}}.

\bibitem{Czakon:2010td}
M.~Czakon, ``{A novel subtraction scheme for double-real radiation at NNLO},''
  \href{http://dx.doi.org/10.1016/j.physletb.2010.08.036}{{\em Phys. Lett. B}
  {\bfseries 693} (2010) 259--268},
  \href{http://arxiv.org/abs/1005.0274}{{\ttfamily arXiv:1005.0274 [hep-ph]}}.

\bibitem{Boughezal:2013uia}
R.~Boughezal, F.~Caola, K.~Melnikov, F.~Petriello, and M.~Schulze, ``{Higgs
  boson production in association with a jet at next-to-next-to-leading order
  in perturbative QCD},'' \href{http://dx.doi.org/10.1007/JHEP06(2013)072}{{\em
  JHEP} {\bfseries 06} (2013) 072},
  \href{http://arxiv.org/abs/1302.6216}{{\ttfamily arXiv:1302.6216 [hep-ph]}}.

\bibitem{Boughezal:2015dra}
R.~Boughezal, F.~Caola, K.~Melnikov, F.~Petriello, and M.~Schulze, ``{Higgs
  boson production in association with a jet at next-to-next-to-leading
  order},'' \href{http://dx.doi.org/10.1103/PhysRevLett.115.082003}{{\em Phys.
  Rev. Lett.} {\bfseries 115} no.~8, (2015) 082003},
  \href{http://arxiv.org/abs/1504.07922}{{\ttfamily arXiv:1504.07922
  [hep-ph]}}.

\bibitem{Caola:2017dug}
F.~Caola, K.~Melnikov, and R.~R\"ontsch, ``{Nested soft-collinear subtractions
  in NNLO QCD computations},''
  \href{http://dx.doi.org/10.1140/epjc/s10052-017-4774-0}{{\em Eur. Phys. J. C}
  {\bfseries 77} no.~4, (2017) 248},
  \href{http://arxiv.org/abs/1702.01352}{{\ttfamily arXiv:1702.01352
  [hep-ph]}}.

\bibitem{Boughezal:2015dva}
R.~Boughezal, C.~Focke, X.~Liu, and F.~Petriello, ``{$W$-boson production in
  association with a jet at next-to-next-to-leading order in perturbative
  QCD},'' \href{http://dx.doi.org/10.1103/PhysRevLett.115.062002}{{\em Phys.
  Rev. Lett.} {\bfseries 115} no.~6, (2015) 062002},
  \href{http://arxiv.org/abs/1504.02131}{{\ttfamily arXiv:1504.02131
  [hep-ph]}}.

\bibitem{Gaunt:2015pea}
J.~Gaunt, M.~Stahlhofen, F.~J. Tackmann, and J.~R. Walsh, ``{N-jettiness
  Subtractions for NNLO QCD Calculations},''
  \href{http://dx.doi.org/10.1007/JHEP09(2015)058}{{\em JHEP} {\bfseries 09}
  (2015) 058}, \href{http://arxiv.org/abs/1505.04794}{{\ttfamily
  arXiv:1505.04794 [hep-ph]}}.

\bibitem{Catani:2007vq}
S.~Catani and M.~Grazzini, ``{An NNLO subtraction formalism in hadron
  collisions and its application to Higgs boson production at the LHC},''
  \href{http://dx.doi.org/10.1103/PhysRevLett.98.222002}{{\em Phys. Rev. Lett.}
  {\bfseries 98} (2007) 222002},
  \href{http://arxiv.org/abs/hep-ph/0703012}{{\ttfamily arXiv:hep-ph/0703012}}.

\bibitem{Cacciari:2015jma}
M.~Cacciari, F.~A. Dreyer, A.~Karlberg, G.~P. Salam, and G.~Zanderighi,
  ``{Fully Differential Vector-Boson-Fusion Higgs Production at
  Next-to-Next-to-Leading Order},''
  \href{http://dx.doi.org/10.1103/PhysRevLett.115.082002}{{\em Phys. Rev.
  Lett.} {\bfseries 115} no.~8, (2015) 082002},
  \href{http://arxiv.org/abs/1506.02660}{{\ttfamily arXiv:1506.02660
  [hep-ph]}}. [Erratum: Phys.Rev.Lett. 120, 139901 (2018)].

\bibitem{Ebert:2020yqt}
M.~A. Ebert, B.~Mistlberger, and G.~Vita, ``{Transverse momentum dependent PDFs
  at N$^3$LO},'' \href{http://dx.doi.org/10.1007/JHEP09(2020)146}{{\em JHEP}
  {\bfseries 09} (2020) 146}, \href{http://arxiv.org/abs/2006.05329}{{\ttfamily
  arXiv:2006.05329 [hep-ph]}}.

\bibitem{Ebert:2020unb}
M.~A. Ebert, B.~Mistlberger, and G.~Vita, ``{$N$-jettiness beam functions at
  N$^{3}$LO},'' \href{http://dx.doi.org/10.1007/JHEP09(2020)143}{{\em JHEP}
  {\bfseries 09} (2020) 143}, \href{http://arxiv.org/abs/2006.03056}{{\ttfamily
  arXiv:2006.03056 [hep-ph]}}.

\bibitem{Chen:2022cvz}
W.~Chen, F.~Feng, Y.~Jia, and X.~Liu, ``{Double-Real-Virtual and
  Double-Virtual-Real Corrections to the Three-Loop Thrust Soft Function},''
  \href{http://arxiv.org/abs/2206.12323}{{\ttfamily arXiv:2206.12323
  [hep-ph]}}.

\bibitem{Baranowski:2022khd}
D.~Baranowski, M.~Delto, K.~Melnikov, and C.-Y. Wang, ``{Same-hemisphere
  three-gluon-emission contribution to the zero-jettiness soft function at N3LO
  QCD},'' \href{http://dx.doi.org/10.1103/PhysRevD.106.014004}{{\em Phys. Rev.
  D} {\bfseries 106} no.~1, (2022) 014004},
  \href{http://arxiv.org/abs/2204.09459}{{\ttfamily arXiv:2204.09459
  [hep-ph]}}.

\bibitem{Cordero:2022gsh}
F.~F. Cordero, A.~von Manteuffel, and T.~Neumann, ``{Computational challenges
  for multi-loop collider phenomenology},'' in {\em {2022 Snowmass Summer
  Study}}.
\newblock 4, 2022.
\newblock \href{http://arxiv.org/abs/2204.04200}{{\ttfamily arXiv:2204.04200
  [hep-ph]}}.

\bibitem{Catani:2011st}
S.~Catani, D.~de~Florian, and G.~Rodrigo, ``{Space-like (versus time-like)
  collinear limits in QCD: Is factorization violated?},''
  \href{http://dx.doi.org/10.1007/JHEP07(2012)026}{{\em JHEP} {\bfseries 07}
  (2012) 026}, \href{http://arxiv.org/abs/1112.4405}{{\ttfamily arXiv:1112.4405
  [hep-ph]}}.

\bibitem{Caola:2021kzt}
F.~Caola, S.~Ferrario~Ravasio, G.~Limatola, K.~Melnikov, and P.~Nason, ``{On
  linear power corrections in certain collider observables},''
  \href{http://dx.doi.org/10.1007/JHEP01(2022)093}{{\em JHEP} {\bfseries 01}
  (2022) 093}, \href{http://arxiv.org/abs/2108.08897}{{\ttfamily
  arXiv:2108.08897 [hep-ph]}}.

\bibitem{Sterman:2022gyf}
G.~Sterman, ``{Comments on collinear factorization},'' in {\em {2022 Snowmass
  Summer Study}}.
\newblock 7, 2022.
\newblock \href{http://arxiv.org/abs/2207.06507}{{\ttfamily arXiv:2207.06507
  [hep-ph]}}.

\bibitem{Sterman:1986aj}
G.~F. Sterman, ``{Summation of Large Corrections to Short Distance Hadronic
  Cross-Sections},'' \href{http://dx.doi.org/10.1016/0550-3213(87)90258-6}{{\em
  Nucl. Phys. B} {\bfseries 281} (1987) 310--364}.

\bibitem{Catani:1989ne}
S.~Catani and L.~Trentadue, ``{Resummation of the QCD Perturbative Series for
  Hard Processes},'' \href{http://dx.doi.org/10.1016/0550-3213(89)90273-3}{{\em
  Nucl. Phys. B} {\bfseries 327} (1989) 323--352}.

\bibitem{Collins:1981uk}
J.~C. Collins and D.~E. Soper, ``{Back-To-Back Jets in QCD},''
  \href{http://dx.doi.org/10.1016/0550-3213(81)90339-4}{{\em Nucl. Phys. B}
  {\bfseries 193} (1981) 381}. [Erratum: Nucl.Phys.B 213, 545 (1983)].

\bibitem{Collins:1984kg}
J.~C. Collins, D.~E. Soper, and G.~F. Sterman, ``{Transverse Momentum
  Distribution in Drell-Yan Pair and W and Z Boson Production},''
  \href{http://dx.doi.org/10.1016/0550-3213(85)90479-1}{{\em Nucl. Phys. B}
  {\bfseries 250} (1985) 199--224}.

\bibitem{Bozzi:2003jy}
G.~Bozzi, S.~Catani, D.~de~Florian, and M.~Grazzini, ``{The q(T) spectrum of
  the Higgs boson at the LHC in QCD perturbation theory},''
  \href{http://dx.doi.org/10.1016/S0370-2693(03)00656-7}{{\em Phys. Lett. B}
  {\bfseries 564} (2003) 65--72},
  \href{http://arxiv.org/abs/hep-ph/0302104}{{\ttfamily arXiv:hep-ph/0302104}}.

\bibitem{Bauer:2000ew}
C.~W. Bauer, S.~Fleming, and M.~E. Luke, ``{Summing Sudakov logarithms in $B
  \to X_s \gamma$ in effective field theory},''
  \href{http://dx.doi.org/10.1103/PhysRevD.63.014006}{{\em Phys. Rev. D}
  {\bfseries 63} (2000) 014006},
  \href{http://arxiv.org/abs/hep-ph/0005275}{{\ttfamily arXiv:hep-ph/0005275}}.

\bibitem{Bauer:2000yr}
C.~W. Bauer, S.~Fleming, D.~Pirjol, and I.~W. Stewart, ``{An Effective field
  theory for collinear and soft gluons: Heavy to light decays},''
  \href{http://dx.doi.org/10.1103/PhysRevD.63.114020}{{\em Phys. Rev. D}
  {\bfseries 63} (2001) 114020},
  \href{http://arxiv.org/abs/hep-ph/0011336}{{\ttfamily arXiv:hep-ph/0011336}}.

\bibitem{Bauer:2001yt}
C.~W. Bauer, D.~Pirjol, and I.~W. Stewart, ``{Soft collinear factorization in
  effective field theory},''
  \href{http://dx.doi.org/10.1103/PhysRevD.65.054022}{{\em Phys. Rev. D}
  {\bfseries 65} (2002) 054022},
  \href{http://arxiv.org/abs/hep-ph/0109045}{{\ttfamily arXiv:hep-ph/0109045}}.

\bibitem{Bauer:2002nz}
C.~W. Bauer, S.~Fleming, D.~Pirjol, I.~Z. Rothstein, and I.~W. Stewart, ``{Hard
  scattering factorization from effective field theory},''
  \href{http://dx.doi.org/10.1103/PhysRevD.66.014017}{{\em Phys. Rev. D}
  {\bfseries 66} (2002) 014017},
  \href{http://arxiv.org/abs/hep-ph/0202088}{{\ttfamily arXiv:hep-ph/0202088}}.

\bibitem{vanBeekveld:2022blq}
M.~van Beekveld {\em et~al.}, ``{Snowmass 2021 White Paper: Resummation for
  future colliders},'' in {\em {2022 Snowmass Summer Study}}.
\newblock 3, 2022.
\newblock \href{http://arxiv.org/abs/2203.07907}{{\ttfamily arXiv:2203.07907
  [hep-ph]}}.

\bibitem{Bahjat-Abbas:2019fqa}
N.~Bahjat-Abbas, D.~Bonocore, J.~Sinninghe~Damst\'e, E.~Laenen, L.~Magnea,
  L.~Vernazza, and C.~D. White, ``{Diagrammatic resummation of
  leading-logarithmic threshold effects at next-to-leading power},''
  \href{http://dx.doi.org/10.1007/JHEP11(2019)002}{{\em JHEP} {\bfseries 11}
  (2019) 002}, \href{http://arxiv.org/abs/1905.13710}{{\ttfamily
  arXiv:1905.13710 [hep-ph]}}.

\bibitem{vanBeekveld:2021hhv}
M.~van Beekveld, E.~Laenen, J.~Sinninghe~Damst\'e, and L.~Vernazza,
  ``{Next-to-leading power threshold corrections for finite order and resummed
  colour-singlet cross sections},''
  \href{http://dx.doi.org/10.1007/JHEP05(2021)114}{{\em JHEP} {\bfseries 05}
  (2021) 114}, \href{http://arxiv.org/abs/2101.07270}{{\ttfamily
  arXiv:2101.07270 [hep-ph]}}.

\bibitem{Beneke:2017ztn}
M.~Beneke, M.~Garny, R.~Szafron, and J.~Wang, ``{Anomalous dimension of
  subleading-power N-jet operators},''
  \href{http://dx.doi.org/10.1007/JHEP03(2018)001}{{\em JHEP} {\bfseries 03}
  (2018) 001}, \href{http://arxiv.org/abs/1712.04416}{{\ttfamily
  arXiv:1712.04416 [hep-ph]}}.

\bibitem{Feige:2017zci}
I.~Feige, D.~W. Kolodrubetz, I.~Moult, and I.~W. Stewart, ``{A Complete Basis
  of Helicity Operators for Subleading Factorization},''
  \href{http://dx.doi.org/10.1007/JHEP11(2017)142}{{\em JHEP} {\bfseries 11}
  (2017) 142}, \href{http://arxiv.org/abs/1703.03411}{{\ttfamily
  arXiv:1703.03411 [hep-ph]}}.

\bibitem{Beneke:2018rbh}
M.~Beneke, M.~Garny, R.~Szafron, and J.~Wang, ``{Anomalous dimension of
  subleading-power $N$-jet operators. Part II},''
  \href{http://dx.doi.org/10.1007/JHEP11(2018)112}{{\em JHEP} {\bfseries 11}
  (2018) 112}, \href{http://arxiv.org/abs/1808.04742}{{\ttfamily
  arXiv:1808.04742 [hep-ph]}}.

\bibitem{Beneke:2019mua}
M.~Beneke, M.~Garny, S.~Jaskiewicz, R.~Szafron, L.~Vernazza, and J.~Wang,
  ``{Leading-logarithmic threshold resummation of Higgs production in gluon
  fusion at next-to-leading power},''
  \href{http://dx.doi.org/10.1007/JHEP01(2020)094}{{\em JHEP} {\bfseries 01}
  (2020) 094}, \href{http://arxiv.org/abs/1910.12685}{{\ttfamily
  arXiv:1910.12685 [hep-ph]}}.

\bibitem{Melnikov:2016emg}
K.~Melnikov and A.~Penin, ``{On the light quark mass effects in Higgs boson
  production in gluon fusion},''
  \href{http://dx.doi.org/10.1007/JHEP05(2016)172}{{\em JHEP} {\bfseries 05}
  (2016) 172}, \href{http://arxiv.org/abs/1602.09020}{{\ttfamily
  arXiv:1602.09020 [hep-ph]}}.

\bibitem{Liu:2018czl}
T.~Liu and A.~Penin, ``{High-Energy Limit of Mass-Suppressed Amplitudes in
  Gauge Theories},'' \href{http://dx.doi.org/10.1007/JHEP11(2018)158}{{\em
  JHEP} {\bfseries 11} (2018) 158},
  \href{http://arxiv.org/abs/1809.04950}{{\ttfamily arXiv:1809.04950
  [hep-ph]}}.

\bibitem{Liu:2019oav}
Z.~L. Liu and M.~Neubert, ``{Factorization at subleading power and
  endpoint-divergent convolutions in $h\to\gamma\gamma$ decay},''
  \href{http://dx.doi.org/10.1007/JHEP04(2020)033}{{\em JHEP} {\bfseries 04}
  (2020) 033}, \href{http://arxiv.org/abs/1912.08818}{{\ttfamily
  arXiv:1912.08818 [hep-ph]}}.

\bibitem{Liu:2020tzd}
Z.~L. Liu, B.~Mecaj, M.~Neubert, and X.~Wang, ``{Factorization at subleading
  power, Sudakov resummation, and endpoint divergences in soft-collinear
  effective theory},''
  \href{http://dx.doi.org/10.1103/PhysRevD.104.014004}{{\em Phys. Rev. D}
  {\bfseries 104} no.~1, (2021) 014004},
  \href{http://arxiv.org/abs/2009.04456}{{\ttfamily arXiv:2009.04456
  [hep-ph]}}.

\bibitem{Kogler:2018hem}
R.~Kogler {\em et~al.}, ``{Jet Substructure at the Large Hadron Collider:
  Experimental Review},''
  \href{http://dx.doi.org/10.1103/RevModPhys.91.045003}{{\em Rev. Mod. Phys.}
  {\bfseries 91} no.~4, (2019) 045003},
  \href{http://arxiv.org/abs/1803.06991}{{\ttfamily arXiv:1803.06991
  [hep-ex]}}.

\bibitem{Marzani:2019hun}
S.~Marzani, G.~Soyez, and M.~Spannowsky, ``{Looking inside jets: an
  introduction to jet substructure and boosted-object phenomenology},''
  \href{http://arxiv.org/abs/1901.10342}{{\ttfamily arXiv:1901.10342
  [hep-ph]}}.

\bibitem{McLerran:1993ni}
L.~D. McLerran and R.~Venugopalan, ``{Computing quark and gluon distribution
  functions for very large nuclei},''
  \href{http://dx.doi.org/10.1103/PhysRevD.49.2233}{{\em Phys. Rev. D}
  {\bfseries 49} (1994) 2233--2241},
  \href{http://arxiv.org/abs/hep-ph/9309289}{{\ttfamily arXiv:hep-ph/9309289}}.

\bibitem{McLerran:1993ka}
L.~D. McLerran and R.~Venugopalan, ``{Gluon distribution functions for very
  large nuclei at small transverse momentum},''
  \href{http://dx.doi.org/10.1103/PhysRevD.49.3352}{{\em Phys. Rev. D}
  {\bfseries 49} (1994) 3352--3355},
  \href{http://arxiv.org/abs/hep-ph/9311205}{{\ttfamily arXiv:hep-ph/9311205}}.

\bibitem{Liu:2020mpy}
H.-Y. Liu, Z.-B. Kang, and X.~Liu, ``{Threshold resummation for hadron
  production in the small-$x$ region},''
  \href{http://dx.doi.org/10.1103/PhysRevD.102.051502}{{\em Phys. Rev. D}
  {\bfseries 102} no.~5, (2020) 051502},
  \href{http://arxiv.org/abs/2004.11990}{{\ttfamily arXiv:2004.11990
  [hep-ph]}}.

\bibitem{Constantinou:2020hdm}
M.~Constantinou {\em et~al.}, ``{Parton distributions and lattice-QCD
  calculations: Toward 3D structure},''
  \href{http://dx.doi.org/10.1016/j.ppnp.2021.103908}{{\em Prog. Part. Nucl.
  Phys.} {\bfseries 121} (2021) 103908},
  \href{http://arxiv.org/abs/2006.08636}{{\ttfamily arXiv:2006.08636
  [hep-ph]}}.

\bibitem{Moch:2017uml}
S.~Moch, B.~Ruijl, T.~Ueda, J.~A.~M. Vermaseren, and A.~Vogt, ``{Four-Loop
  Non-Singlet Splitting Functions in the Planar Limit and Beyond},''
  \href{http://dx.doi.org/10.1007/JHEP10(2017)041}{{\em JHEP} {\bfseries 10}
  (2017) 041}, \href{http://arxiv.org/abs/1707.08315}{{\ttfamily
  arXiv:1707.08315 [hep-ph]}}.

\bibitem{Moch:2018wjh}
S.~Moch, B.~Ruijl, T.~Ueda, J.~A.~M. Vermaseren, and A.~Vogt, ``{On quartic
  colour factors in splitting functions and the gluon cusp anomalous
  dimension},'' \href{http://dx.doi.org/10.1016/j.physletb.2018.06.017}{{\em
  Phys. Lett. B} {\bfseries 782} (2018) 627--632},
  \href{http://arxiv.org/abs/1805.09638}{{\ttfamily arXiv:1805.09638
  [hep-ph]}}.

\bibitem{Moch:2021qrk}
S.~Moch, B.~Ruijl, T.~Ueda, J.~A.~M. Vermaseren, and A.~Vogt, ``{Low moments of
  the four-loop splitting functions in QCD},''
  \href{http://dx.doi.org/10.1016/j.physletb.2021.136853}{{\em Phys. Lett. B}
  {\bfseries 825} (2022) 136853},
  \href{http://arxiv.org/abs/2111.15561}{{\ttfamily arXiv:2111.15561
  [hep-ph]}}.

\bibitem{Blumlein:2006be}
J.~Blumlein, H.~Bottcher, and A.~Guffanti, ``{Non-singlet QCD analysis of deep
  inelastic world data at O(alpha(s)**3)},''
  \href{http://dx.doi.org/10.1016/j.nuclphysb.2007.03.035}{{\em Nucl. Phys. B}
  {\bfseries 774} (2007) 182--207},
  \href{http://arxiv.org/abs/hep-ph/0607200}{{\ttfamily arXiv:hep-ph/0607200}}.

\bibitem{Blumlein:2021lmf}
J.~Bl\"umlein and M.~Saragnese, ``{The N3LO scheme-invariant QCD evolution of
  the non-singlet structure functions F2NS(x,Q2) and g1NS(x,Q2)},''
  \href{http://dx.doi.org/10.1016/j.physletb.2021.136589}{{\em Phys. Lett. B}
  {\bfseries 820} (2021) 136589},
  \href{http://arxiv.org/abs/2107.01293}{{\ttfamily arXiv:2107.01293
  [hep-ph]}}.

\bibitem{Blumlein:2022ndg}
J.~Bl\"umlein, P.~Marquard, C.~Schneider, and K.~Sch\"onwald, ``{The Two-Loop
  Massless Off-Shell QCD Operator Matrix Elements to Finite Terms},''
  \href{http://arxiv.org/abs/2202.03216}{{\ttfamily arXiv:2202.03216
  [hep-ph]}}.

\bibitem{NNPDF:2019vjt}
{\bfseries NNPDF} Collaboration, R.~Abdul~Khalek {\em et~al.}, ``{A first
  determination of parton distributions with theoretical uncertainties},''
  \href{http://dx.doi.org/10.1140/epjc/s10052-019-7364-5}{{\em Eur. Phys. J.}
  {\bfseries C} (2019) 79:838},
  \href{http://arxiv.org/abs/1905.04311}{{\ttfamily arXiv:1905.04311
  [hep-ph]}}.

\bibitem{NNPDF:2019ubu}
{\bfseries NNPDF} Collaboration, R.~Abdul~Khalek {\em et~al.}, ``{Parton
  Distributions with Theory Uncertainties: General Formalism and First
  Phenomenological Studies},''
  \href{http://dx.doi.org/10.1140/epjc/s10052-019-7401-4}{{\em Eur. Phys. J. C}
  {\bfseries 79} no.~11, (2019) 931},
  \href{http://arxiv.org/abs/1906.10698}{{\ttfamily arXiv:1906.10698
  [hep-ph]}}.

\bibitem{AbdulKhalek:2021gbh}
R.~Abdul~Khalek {\em et~al.}, ``{Science Requirements and Detector Concepts for
  the Electron-Ion Collider: EIC Yellow Report},''
  \href{http://arxiv.org/abs/2103.05419}{{\ttfamily arXiv:2103.05419
  [physics.ins-det]}}.

\bibitem{Khalek:2021ulf}
R.~A. Khalek, J.~J. Ethier, E.~R. Nocera, and J.~Rojo, ``{Self-consistent
  determination of proton and nuclear PDFs at the Electron Ion Collider},''
  \href{http://dx.doi.org/10.1103/PhysRevD.103.096005}{{\em Phys. Rev. D}
  {\bfseries 103} no.~9, (2021) 096005},
  \href{http://arxiv.org/abs/2102.00018}{{\ttfamily arXiv:2102.00018
  [hep-ph]}}.

\bibitem{Bahr:2008pv}
M.~Bahr {\em et~al.}, ``{Herwig++ Physics and Manual},''
  \href{http://dx.doi.org/10.1140/epjc/s10052-008-0798-9}{{\em Eur. Phys. J. C}
  {\bfseries 58} (2008) 639--707},
  \href{http://arxiv.org/abs/0803.0883}{{\ttfamily arXiv:0803.0883 [hep-ph]}}.

\bibitem{Sjostrand:2014zea}
T.~Sj\"ostrand, S.~Ask, J.~R. Christiansen, R.~Corke, N.~Desai, P.~Ilten,
  S.~Mrenna, S.~Prestel, C.~O. Rasmussen, and P.~Z. Skands, ``{An introduction
  to PYTHIA 8.2}'' \href{http://dx.doi.org/10.1016/j.cpc.2015.01.024}{{\em
  Comput. Phys. Commun.} {\bfseries 191} (2015) 159--177},
  \href{http://arxiv.org/abs/1410.3012}{{\ttfamily arXiv:1410.3012 [hep-ph]}}.

\bibitem{Gleisberg:2008ta}
T.~Gleisberg, S.~Hoeche, F.~Krauss, M.~Schonherr, S.~Schumann, F.~Siegert, and
  J.~Winter, ``{Event generation with SHERPA 1.1}''
  \href{http://dx.doi.org/10.1088/1126-6708/2009/02/007}{{\em JHEP} {\bfseries
  02} (2009) 007}, \href{http://arxiv.org/abs/0811.4622}{{\ttfamily
  arXiv:0811.4622 [hep-ph]}}.

\bibitem{Darvishi:2022gqt}
N.~Darvishi, J.~I. M.~R. Masouminia, Z.~Nagy, P.~Richardson, and D.~E. Soper,
  ``{Future prospects for parton showers},'' in {\em {2022 Snowmass Summer
  Study}}.
\newblock 3, 2022.
\newblock \href{http://arxiv.org/abs/2203.06799}{{\ttfamily arXiv:2203.06799
  [hep-ph]}}.

\bibitem{Campbell:2022qmc}
J.~M. Campbell {\em et~al.}, ``{Event Generators for High-Energy Physics
  Experiments},'' in {\em {2022 Snowmass Summer Study}}.
\newblock 3, 2022.
\newblock \href{http://arxiv.org/abs/2203.11110}{{\ttfamily arXiv:2203.11110
  [hep-ph]}}.

\bibitem{Feruglio:1992wf}
F.~Feruglio, ``{The Chiral approach to the electroweak interactions},''
  \href{http://dx.doi.org/10.1142/S0217751X93001946}{{\em Int. J. Mod. Phys. A}
  {\bfseries 8} (1993) 4937--4972},
  \href{http://arxiv.org/abs/hep-ph/9301281}{{\ttfamily arXiv:hep-ph/9301281}}.

\bibitem{Alonso:2012px}
R.~Alonso, M.~B. Gavela, L.~Merlo, S.~Rigolin, and J.~Yepes, ``{The Effective
  Chiral Lagrangian for a Light Dynamical ''Higgs Particle''},''
  \href{http://dx.doi.org/10.1016/j.physletb.2013.04.037}{{\em Phys. Lett. B}
  {\bfseries 722} (2013) 330--335},
  \href{http://arxiv.org/abs/1212.3305}{{\ttfamily arXiv:1212.3305 [hep-ph]}}.
  [Erratum: Phys.Lett.B 726, 926 (2013)].

\bibitem{Brivio:2016fzo}
I.~Brivio, J.~Gonzalez-Fraile, M.~C. Gonzalez-Garcia, and L.~Merlo, ``{The
  complete HEFT Lagrangian after the LHC Run I},''
  \href{http://dx.doi.org/10.1140/epjc/s10052-016-4211-9}{{\em Eur. Phys. J. C}
  {\bfseries 76} no.~7, (2016) 416},
  \href{http://arxiv.org/abs/1604.06801}{{\ttfamily arXiv:1604.06801
  [hep-ph]}}.

\bibitem{Buchmuller:1985jz}
W.~Buchmuller and D.~Wyler, ``{Effective Lagrangian Analysis of New
  Interactions and Flavor Conservation},''
  \href{http://dx.doi.org/10.1016/0550-3213(86)90262-2}{{\em Nucl. Phys. B}
  {\bfseries 268} (1986) 621--653}.

\bibitem{Arzt:1994gp}
C.~Arzt, M.~B. Einhorn, and J.~Wudka, ``{Patterns of deviation from the
  standard model},'' \href{http://dx.doi.org/10.1016/0550-3213(94)00336-D}{{\em
  Nucl. Phys. B} {\bfseries 433} (1995) 41--66},
  \href{http://arxiv.org/abs/hep-ph/9405214}{{\ttfamily arXiv:hep-ph/9405214}}.

\bibitem{Grzadkowski:2010es}
B.~Grzadkowski, M.~Iskrzynski, M.~Misiak, and J.~Rosiek, ``{Dimension-Six Terms
  in the Standard Model Lagrangian},''
  \href{http://dx.doi.org/10.1007/JHEP10(2010)085}{{\em JHEP} {\bfseries 10}
  (2010) 085}, \href{http://arxiv.org/abs/1008.4884}{{\ttfamily arXiv:1008.4884
  [hep-ph]}}.

\bibitem{Alioli:2022fng}
S.~Alioli {\em et~al.}, ``{Theoretical developments in the SMEFT at dimension-8
  and beyond},'' in {\em {2022 Snowmass Summer Study}}.
\newblock 3, 2022.
\newblock \href{http://arxiv.org/abs/2203.06771}{{\ttfamily arXiv:2203.06771
  [hep-ph]}}.

\bibitem{Lehman:2015via}
L.~Lehman and A.~Martin, ``{Hilbert Series for Constructing Lagrangians:
  expanding the phenomenologist's toolbox},''
  \href{http://dx.doi.org/10.1103/PhysRevD.91.105014}{{\em Phys. Rev. D}
  {\bfseries 91} (2015) 105014},
  \href{http://arxiv.org/abs/1503.07537}{{\ttfamily arXiv:1503.07537
  [hep-ph]}}.

\bibitem{Lehman:2015coa}
L.~Lehman and A.~Martin, ``{Low-derivative operators of the Standard Model
  effective field theory via Hilbert series methods},''
  \href{http://dx.doi.org/10.1007/JHEP02(2016)081}{{\em JHEP} {\bfseries 02}
  (2016) 081}, \href{http://arxiv.org/abs/1510.00372}{{\ttfamily
  arXiv:1510.00372 [hep-ph]}}.

\bibitem{Henning:2015daa}
B.~Henning, X.~Lu, T.~Melia, and H.~Murayama, ``{Hilbert series and operator
  bases with derivatives in effective field theories},''
  \href{http://dx.doi.org/10.1007/s00220-015-2518-2}{{\em Commun. Math. Phys.}
  {\bfseries 347} no.~2, (2016) 363--388},
  \href{http://arxiv.org/abs/1507.07240}{{\ttfamily arXiv:1507.07240
  [hep-th]}}.

\bibitem{Henning:2015alf}
B.~Henning, X.~Lu, T.~Melia, and H.~Murayama, ``{2, 84, 30, 993, 560, 15456,
  11962, 261485, ...: Higher dimension operators in the SM EFT},''
  \href{http://dx.doi.org/10.1007/JHEP08(2017)016}{{\em JHEP} {\bfseries 08}
  (2017) 016}, \href{http://arxiv.org/abs/1512.03433}{{\ttfamily
  arXiv:1512.03433 [hep-ph]}}. [Erratum: JHEP 09, 019 (2019)].

\bibitem{Murphy:2020rsh}
C.~W. Murphy, ``{Dimension-8 operators in the Standard Model Eective Field
  Theory},'' \href{http://dx.doi.org/10.1007/JHEP10(2020)174}{{\em JHEP}
  {\bfseries 10} (2020) 174}, \href{http://arxiv.org/abs/2005.00059}{{\ttfamily
  arXiv:2005.00059 [hep-ph]}}.

\bibitem{Li:2020gnx}
H.-L. Li, Z.~Ren, J.~Shu, M.-L. Xiao, J.-H. Yu, and Y.-H. Zheng, ``{Complete
  set of dimension-eight operators in the standard model effective field
  theory},'' \href{http://dx.doi.org/10.1103/PhysRevD.104.015026}{{\em Phys.
  Rev. D} {\bfseries 104} no.~1, (2021) 015026},
  \href{http://arxiv.org/abs/2005.00008}{{\ttfamily arXiv:2005.00008
  [hep-ph]}}.

\bibitem{Helset:2020yio}
A.~Helset, A.~Martin, and M.~Trott, ``{The Geometric Standard Model Effective
  Field Theory},'' \href{http://dx.doi.org/10.1007/JHEP03(2020)163}{{\em JHEP}
  {\bfseries 03} (2020) 163}, \href{http://arxiv.org/abs/2001.01453}{{\ttfamily
  arXiv:2001.01453 [hep-ph]}}.

\bibitem{Durieux:2019eor}
G.~Durieux, T.~Kitahara, Y.~Shadmi, and Y.~Weiss, ``{The electroweak effective
  field theory from on-shell amplitudes},''
  \href{http://dx.doi.org/10.1007/JHEP01(2020)119}{{\em JHEP} {\bfseries 01}
  (2020) 119}, \href{http://arxiv.org/abs/1909.10551}{{\ttfamily
  arXiv:1909.10551 [hep-ph]}}.

\bibitem{Martin:2021vwf}
A.~Martin and M.~Trott, ``{The $ggh$ variations},''
  \href{http://arxiv.org/abs/2109.05595}{{\ttfamily arXiv:2109.05595
  [hep-ph]}}.

\bibitem{Jenkins:2013zja}
E.~E. Jenkins, A.~V. Manohar, and M.~Trott, ``{Renormalization Group Evolution
  of the Standard Model Dimension Six Operators I: Formalism and lambda
  Dependence},'' \href{http://dx.doi.org/10.1007/JHEP10(2013)087}{{\em JHEP}
  {\bfseries 10} (2013) 087}, \href{http://arxiv.org/abs/1308.2627}{{\ttfamily
  arXiv:1308.2627 [hep-ph]}}.

\bibitem{Jenkins:2013wua}
E.~E. Jenkins, A.~V. Manohar, and M.~Trott, ``{Renormalization Group Evolution
  of the Standard Model Dimension Six Operators II: Yukawa Dependence},''
  \href{http://dx.doi.org/10.1007/JHEP01(2014)035}{{\em JHEP} {\bfseries 01}
  (2014) 035}, \href{http://arxiv.org/abs/1310.4838}{{\ttfamily arXiv:1310.4838
  [hep-ph]}}.

\bibitem{Alonso:2013hga}
R.~Alonso, E.~E. Jenkins, A.~V. Manohar, and M.~Trott, ``{Renormalization Group
  Evolution of the Standard Model Dimension Six Operators III: Gauge Coupling
  Dependence and Phenomenology},''
  \href{http://dx.doi.org/10.1007/JHEP04(2014)159}{{\em JHEP} {\bfseries 04}
  (2014) 159}, \href{http://arxiv.org/abs/1312.2014}{{\ttfamily arXiv:1312.2014
  [hep-ph]}}.

\bibitem{Chala:2021pll}
M.~Chala, G.~Guedes, M.~Ramos, and J.~Santiago, ``{Towards the renormalisation
  of the Standard Model effective field theory to dimension eight: Bosonic
  interactions I},''
  \href{http://dx.doi.org/10.21468/SciPostPhys.11.3.065}{{\em SciPost Phys.}
  {\bfseries 11} (2021) 065}, \href{http://arxiv.org/abs/2106.05291}{{\ttfamily
  arXiv:2106.05291 [hep-ph]}}.

\bibitem{Chala:2021wpj}
M.~Chala and J.~Santiago, ``{Positivity bounds in the Standard Model effective
  field theory beyond tree level},''
  \href{http://arxiv.org/abs/2110.01624}{{\ttfamily arXiv:2110.01624
  [hep-ph]}}.

\bibitem{Corbett:2021eux}
T.~Corbett, A.~Helset, A.~Martin, and M.~Trott, ``{EWPD in the SMEFT to
  dimension eight},'' \href{http://dx.doi.org/10.1007/JHEP06(2021)076}{{\em
  JHEP} {\bfseries 06} (2021) 076},
  \href{http://arxiv.org/abs/2102.02819}{{\ttfamily arXiv:2102.02819
  [hep-ph]}}.

\bibitem{Alioli:2020kez}
S.~Alioli, R.~Boughezal, E.~Mereghetti, and F.~Petriello, ``{Novel angular
  dependence in Drell-Yan lepton production via dimension-8 operators},''
  \href{http://dx.doi.org/10.1016/j.physletb.2020.135703}{{\em Phys. Lett. B}
  {\bfseries 809} (2020) 135703},
  \href{http://arxiv.org/abs/2003.11615}{{\ttfamily arXiv:2003.11615
  [hep-ph]}}.

\bibitem{Boughezal:2022nof}
R.~Boughezal, Y.~Huang, and F.~Petriello, ``{Exploring the SMEFT at dimension-8
  with Drell-Yan transverse momentum measurements},''
  \href{http://arxiv.org/abs/2207.01703}{{\ttfamily arXiv:2207.01703
  [hep-ph]}}.

\bibitem{Boughezal:2021tih}
R.~Boughezal, E.~Mereghetti, and F.~Petriello, ``{Dilepton production in the
  SMEFT at O(1/\ensuremath{\Lambda}4)},''
  \href{http://dx.doi.org/10.1103/PhysRevD.104.095022}{{\em Phys. Rev. D}
  {\bfseries 104} no.~9, (2021) 095022},
  \href{http://arxiv.org/abs/2106.05337}{{\ttfamily arXiv:2106.05337
  [hep-ph]}}.

\bibitem{Mereghetti:2013bta}
E.~Mereghetti, J.~de~Vries, R.~G.~E. Timmermans, and U.~van Kolck, ``{Toroidal
  Quadrupole Form Factor of the Deuteron},''
  \href{http://dx.doi.org/10.1103/PhysRevC.88.034001}{{\em Phys. Rev. C}
  {\bfseries 88} (2013) 034001},
  \href{http://arxiv.org/abs/1305.7049}{{\ttfamily arXiv:1305.7049 [hep-ph]}}.

\bibitem{Boughezal:2021kla}
R.~Boughezal, F.~Petriello, and D.~Wiegand, ``{Disentangling Standard Model EFT
  operators with future low-energy parity-violating electron scattering
  experiments},'' \href{http://dx.doi.org/10.1103/PhysRevD.104.016005}{{\em
  Phys. Rev. D} {\bfseries 104} no.~1, (2021) 016005},
  \href{http://arxiv.org/abs/2104.03979}{{\ttfamily arXiv:2104.03979
  [hep-ph]}}.

\bibitem{Dawson:2022cmu}
S.~Dawson, D.~Fontes, S.~Homiller, and M.~Sullivan, ``{Beyond 6: the role of
  dimension-8 operators in an EFT for the 2HDM},''
  \href{http://arxiv.org/abs/2205.01561}{{\ttfamily arXiv:2205.01561
  [hep-ph]}}.

\bibitem{Cohen:2020qvb}
T.~Cohen, X.~Lu, and Z.~Zhang, ``{STrEAMlining EFT Matching},''
  \href{http://dx.doi.org/10.21468/SciPostPhys.10.5.098}{{\em SciPost Phys.}
  {\bfseries 10} no.~5, (2021) 098},
  \href{http://arxiv.org/abs/2012.07851}{{\ttfamily arXiv:2012.07851
  [hep-ph]}}.

\bibitem{Kobayashi:1973fv}
M.~Kobayashi and T.~Maskawa, ``{CP Violation in the Renormalizable Theory of
  Weak Interaction},'' \href{http://dx.doi.org/10.1143/PTP.49.652}{{\em Prog.
  Theor. Phys.} {\bfseries 49} (1973) 652--657}.

\bibitem{Cabibbo:1963yz}
N.~Cabibbo, ``{Unitary Symmetry and Leptonic Decays},''
  \href{http://dx.doi.org/10.1103/PhysRevLett.10.531}{{\em Phys. Rev. Lett.}
  {\bfseries 10} (1963) 531--533}.

\bibitem{Glashow:1970gm}
S.~L. Glashow, J.~Iliopoulos, and L.~Maiani, ``{Weak Interactions with
  Lepton-Hadron Symmetry},''
  \href{http://dx.doi.org/10.1103/PhysRevD.2.1285}{{\em Phys. Rev. D}
  {\bfseries 2} (1970) 1285--1292}.

\bibitem{Gaillard:1974hs}
M.~K. Gaillard and B.~W. Lee, ``{Rare Decay Modes of the K-Mesons in Gauge
  Theories},'' \href{http://dx.doi.org/10.1103/PhysRevD.10.897}{{\em Phys. Rev.
  D} {\bfseries 10} (1974) 897}.

\bibitem{Vainshtein:1973md}
A.~I. Vainshtein and I.~B. Khriplovich, ``{Restrictions on masses of
  supercharged hadrons in the weinberg model},'' {\em Pisma Zh. Eksp. Teor.
  Fiz.} {\bfseries 18} (1973) 141--145.

\bibitem{Ligeti:2015kwa}
Z.~Ligeti, \href{http://dx.doi.org/10.1142/9789814678766_0006}{``{TASI Lectures
  on Flavor Physics},''} in {\em {Theoretical Advanced Study Institute in
  Elementary Particle Physics}: {Journeys Through the Precision Frontier:
  Amplitudes for Colliders}}, pp.~297--340.
\newblock 2015.
\newblock \href{http://arxiv.org/abs/1502.01372}{{\ttfamily arXiv:1502.01372
  [hep-ph]}}.

\bibitem{Grossman:2017thq}
Y.~Grossman and P.~Tanedo,
  \href{http://dx.doi.org/10.1142/9789813233348_0004}{``{Just a Taste: Lectures
  on Flavor Physics},''} in {\em {Theoretical Advanced Study Institute in
  Elementary Particle Physics}: {Anticipating the Next Discoveries in Particle
  Physics}}, pp.~109--295.
\newblock 2018.
\newblock \href{http://arxiv.org/abs/1711.03624}{{\ttfamily arXiv:1711.03624
  [hep-ph]}}.

\bibitem{Gori:2019ybw}
S.~Gori, ``{TASI lectures on flavor physics},'' {\em PoS} {\bfseries TASI2018}
  (2019) 013. \url{https://inspirehep.net/literature/1746427}.

\bibitem{Grossman:2021xfq}
Y.~Grossman and Z.~Ligeti, ``{Theoretical challenges for flavor physics},''
  \href{http://dx.doi.org/10.1140/epjp/s13360-021-01845-7}{{\em Eur. Phys. J.
  Plus} {\bfseries 136} no.~9, (2021) 912},
  \href{http://arxiv.org/abs/2106.12168}{{\ttfamily arXiv:2106.12168
  [hep-ph]}}.

\bibitem{Bediaga:2018lhg}
{\bfseries LHCb} Collaboration, R.~Aaij {\em et~al.}, ``{Physics case for an
  LHCb Upgrade II -- Opportunities in flavour physics, and beyond, in the
  HL-LHC era},'' \href{http://arxiv.org/abs/1808.08865}{{\ttfamily
  arXiv:1808.08865 [hep-ex]}}.

\bibitem{Cerri:2018ypt}
A.~Cerri {\em et~al.}, ``{Report from Working Group 4}: {Opportunities in
  Flavour Physics at the HL-LHC and HE-LHC},''
  \href{http://dx.doi.org/10.23731/CYRM-2019-007.867}{{\em CERN Yellow Rep.
  Monogr.} {\bfseries 7} (2019) 867--1158},
  \href{http://arxiv.org/abs/1812.07638}{{\ttfamily arXiv:1812.07638
  [hep-ph]}}.

\bibitem{Belle-II:2018jsg}
{\bfseries Belle~II} Collaboration, W.~Altmannshofer {\em et~al.}, ``{The Belle
  II Physics Book},'' \href{http://dx.doi.org/10.1093/ptep/ptz106}{{\em PTEP}
  {\bfseries 2019} no.~12, (2019) 123C01},
  \href{http://arxiv.org/abs/1808.10567}{{\ttfamily arXiv:1808.10567
  [hep-ex]}}. [Erratum: PTEP 2020, 029201 (2020)].

\bibitem{BESIII:2022mxl}
{\bfseries BESIII} Collaboration, H.~B. Li {\em et~al.}, ``{Physics in the
  $\tau$-charm Region at BESIII},'' in {\em {2022 Snowmass Summer Study}}.
\newblock 4, 2022.
\newblock \href{http://arxiv.org/abs/2204.08943}{{\ttfamily arXiv:2204.08943
  [hep-ex]}}.

\bibitem{Charles:2004jd}
J.~Charles, A.~Hocker, H.~Lacker, S.~Laplace, F.~R. Le~Diberder, J.~Malcles,
  J.~Ocariz, M.~Pivk, and L.~Roos, ``{CP violation and the CKM matrix:
  Assessing the impact of the asymmetric $B$ factories},''
  \href{http://dx.doi.org/10.1140/epjc/s2005-02169-1}{{\em Eur. Phys. J. C}
  {\bfseries 41} no.~1, (2005) 1--131},
  \href{http://arxiv.org/abs/hep-ph/0406184}{{\ttfamily arXiv:hep-ph/0406184}}.
  and updates at \url{http://ckmfitter.in2p3.fr/}.

\bibitem{Charles:2020dfl}
J.~Charles, S.~Descotes-Genon, Z.~Ligeti, S.~Monteil, M.~Papucci, K.~Trabelsi,
  and L.~Vale~Silva, ``{New physics in $B$ meson mixing: future sensitivity and
  limitations},'' \href{http://dx.doi.org/10.1103/PhysRevD.102.056023}{{\em
  Phys. Rev. D} {\bfseries 102} no.~5, (2020) 056023},
  \href{http://arxiv.org/abs/2006.04824}{{\ttfamily arXiv:2006.04824
  [hep-ph]}}.

\bibitem{EuropeanStrategyforParticlePhysicsPreparatoryGroup:2019qin}
R.~K. Ellis {\em et~al.}, ``{Physics Briefing Book}: {Input for the European
  Strategy for Particle Physics Update 2020},''
  \href{http://arxiv.org/abs/1910.11775}{{\ttfamily arXiv:1910.11775
  [hep-ex]}}.

\bibitem{Dine:1995ag}
M.~Dine, A.~E. Nelson, Y.~Nir, and Y.~Shirman, ``{New tools for low-energy
  dynamical supersymmetry breaking},''
  \href{http://dx.doi.org/10.1103/PhysRevD.53.2658}{{\em Phys. Rev. D}
  {\bfseries 53} (1996) 2658--2669},
  \href{http://arxiv.org/abs/hep-ph/9507378}{{\ttfamily arXiv:hep-ph/9507378}}.

\bibitem{Nir:1993mx}
Y.~Nir and N.~Seiberg, ``{Should squarks be degenerate?},''
  \href{http://dx.doi.org/10.1016/0370-2693(93)90942-B}{{\em Phys. Lett. B}
  {\bfseries 309} (1993) 337--343},
  \href{http://arxiv.org/abs/hep-ph/9304307}{{\ttfamily arXiv:hep-ph/9304307}}.

\bibitem{Cohen:1996vb}
A.~G. Cohen, D.~B. Kaplan, and A.~E. Nelson, ``{The More minimal supersymmetric
  standard model},''
  \href{http://dx.doi.org/10.1016/S0370-2693(96)01183-5}{{\em Phys. Lett.}
  {\bfseries B388} (1996) 588--598},
\href{http://arxiv.org/abs/hep-ph/9607394}{{\ttfamily arXiv:hep-ph/9607394
  [hep-ph]}}.
%%CITATION = HEP-PH/9607394;%%.

\bibitem{Kribs:2007ac}
G.~D. Kribs, E.~Poppitz, and N.~Weiner, ``{Flavor in supersymmetry with an
  extended R-symmetry},''
  \href{http://dx.doi.org/10.1103/PhysRevD.78.055010}{{\em Phys. Rev. D}
  {\bfseries 78} (2008) 055010},
  \href{http://arxiv.org/abs/0712.2039}{{\ttfamily arXiv:0712.2039 [hep-ph]}}.

\bibitem{Arkani-Hamed:2005zuc}
N.~Arkani-Hamed, S.~Dimopoulos, and S.~Kachru, ``{Predictive landscapes and new
  physics at a TeV},'' \href{http://arxiv.org/abs/hep-th/0501082}{{\ttfamily
  arXiv:hep-th/0501082}}.

\bibitem{Fox:2005yp}
P.~J. Fox, D.~E. Kaplan, E.~Katz, E.~Poppitz, V.~Sanz, M.~Schmaltz, M.~D.
  Schwartz, and N.~Weiner, ``{Supersplit supersymmetry},''
  \href{http://arxiv.org/abs/hep-th/0503249}{{\ttfamily arXiv:hep-th/0503249}}.

\bibitem{Arkani-Hamed:1999ylh}
N.~Arkani-Hamed and M.~Schmaltz, ``{Hierarchies without symmetries from extra
  dimensions},'' \href{http://dx.doi.org/10.1103/PhysRevD.61.033005}{{\em Phys.
  Rev. D} {\bfseries 61} (2000) 033005},
  \href{http://arxiv.org/abs/hep-ph/9903417}{{\ttfamily arXiv:hep-ph/9903417}}.

\bibitem{Chivukula:1987py}
R.~S. Chivukula and H.~Georgi, ``{Composite Technicolor Standard Model},''
  \href{http://dx.doi.org/10.1016/0370-2693(87)90713-1}{{\em Phys. Lett. B}
  {\bfseries 188} (1987) 99--104}.

\bibitem{Hall:1990ac}
L.~J. Hall and L.~Randall, ``{Weak scale effective supersymmetry},''
  \href{http://dx.doi.org/10.1103/PhysRevLett.65.2939}{{\em Phys. Rev. Lett.}
  {\bfseries 65} (1990) 2939--2942}.

\bibitem{DAmbrosio:2002vsn}
G.~D'Ambrosio, G.~F. Giudice, G.~Isidori, and A.~Strumia, ``{Minimal flavor
  violation: An Effective field theory approach},''
  \href{http://dx.doi.org/10.1016/S0550-3213(02)00836-2}{{\em Nucl. Phys. B}
  {\bfseries 645} (2002) 155--187},
  \href{http://arxiv.org/abs/hep-ph/0207036}{{\ttfamily arXiv:hep-ph/0207036}}.

\bibitem{Altmannshofer:2022aml}
W.~Altmannshofer and J.~Zupan, ``{Snowmass White Paper: Flavor Model
  Building},'' in {\em {2022 Snowmass Summer Study}}.
\newblock 3, 2022.
\newblock \href{http://arxiv.org/abs/2203.07726}{{\ttfamily arXiv:2203.07726
  [hep-ph]}}.

\bibitem{Buras:2011we}
A.~J. Buras, ``{Climbing NLO and NNLO Summits of Weak Decays},''
  \href{http://arxiv.org/abs/1102.5650}{{\ttfamily arXiv:1102.5650 [hep-ph]}}.

\bibitem{Bertolini:1986th}
S.~Bertolini, F.~Borzumati, and A.~Masiero, ``{QCD Enhancement of Radiative b
  Decays},''
\href{http://dx.doi.org/10.1103/PhysRevLett.59.180}{{\em Phys. Rev. Lett.}
  {\bfseries 59} (1987) 180}.
%%CITATION = PRLTA,59,180;%%.

\bibitem{Grinstein:1987vj}
B.~Grinstein, R.~P. Springer, and M.~B. Wise, ``{Effective Hamiltonian for Weak
  Radiative $B$ Meson Decay},''
\href{http://dx.doi.org/10.1016/0370-2693(88)90868-4}{{\em Phys.~Lett.~B}
  {\bfseries 202} (1988) 138--144}.
%%CITATION = PHLTA,B202,138;%%.

\bibitem{Hou:1987kf}
W.-S. Hou and R.~S. Willey, ``{Effects of Charged Higgs Bosons on the Processes
  $b \to s\gamma$, $b \to s g^*$ and $b \to s \ell^+\ell^-$},''
\href{http://dx.doi.org/10.1016/0370-2693(88)91870-9}{{\em Phys.~Lett.~B}
  {\bfseries 202} (1988) 591--595}.
%%CITATION = PHLTA,B202,591;%%.

\bibitem{Grinstein:1987pu}
B.~Grinstein and M.~B. Wise, ``{Weak Radiative $B$ Meson Decay as a Probe of
  the Higgs Sector},''
\href{http://dx.doi.org/10.1016/0370-2693(88)90227-4}{{\em Phys.~Lett.~B}
  {\bfseries 201} (1988) 274--278}.
%%CITATION = PHLTA,B201,274;%%.

\bibitem{Misiak:2006zs}
M.~Misiak {\em et~al.}, ``{Estimate of $\mathcal{B} (\bar B \to X_s \gamma)$ at
  $O(\alpha_s^2)$},''
  \href{http://dx.doi.org/10.1103/PhysRevLett.98.022002}{{\em Phys. Rev. Lett.}
  {\bfseries 98} (2007) 022002},
  \href{http://arxiv.org/abs/hep-ph/0609232}{{\ttfamily arXiv:hep-ph/0609232}}.

\bibitem{Misiak:2015xwa}
M.~Misiak {\em et~al.}, ``{Updated NNLO QCD predictions for the weak radiative
  $B$-meson decays},''
  \href{http://dx.doi.org/10.1103/PhysRevLett.114.221801}{{\em Phys. Rev.
  Lett.} {\bfseries 114} no.~22, (2015) 221801},
  \href{http://arxiv.org/abs/1503.01789}{{\ttfamily arXiv:1503.01789
  [hep-ph]}}.

\bibitem{Neubert:1993um}
M.~Neubert, ``{Analysis of the photon spectrum in inclusive $B \to X_s\gamma$
  decays},'' \href{http://dx.doi.org/10.1103/PhysRevD.49.4623}{{\em
  Phys.~Rev.~D} {\bfseries 49} (1994) 4623--4633},
\href{http://arxiv.org/abs/hep-ph/9312311}{{\ttfamily hep-ph/9312311}}.
%%CITATION = HEP-PH/9312311;%%.

\bibitem{Bigi:1993ex}
I.~I.~Y. Bigi, M.~A. Shifman, N.~G. Uraltsev, and A.~I. Vainshtein, ``{On the
  motion of heavy quarks inside hadrons: Universal distributions and inclusive
  decays},'' \href{http://dx.doi.org/10.1142/S0217751X94000996}{{\em Int. J.
  Mod. Phys.} {\bfseries A09} (1994) 2467--2504},
\href{http://arxiv.org/abs/hep-ph/9312359}{{\ttfamily hep-ph/9312359}}.
%%CITATION = HEP-PH/9312359;%%.

\bibitem{Ligeti:2008ac}
Z.~Ligeti, I.~W. Stewart, and F.~J. Tackmann, ``{Treating the $b$ quark
  distribution function with reliable uncertainties},''
  \href{http://dx.doi.org/10.1103/PhysRevD.78.114014}{{\em Phys. Rev. D}
  {\bfseries 78} (2008) 114014},
  \href{http://arxiv.org/abs/0807.1926}{{\ttfamily arXiv:0807.1926 [hep-ph]}}.

\bibitem{Bernlochner:2020jlt}
{\bfseries SIMBA} Collaboration, F.~U. Bernlochner, H.~Lacker, Z.~Ligeti, I.~W.
  Stewart, F.~J. Tackmann, and K.~Tackmann, ``{Precision Global Determination
  of the B\textrightarrow{}Xs\ensuremath{\gamma} Decay Rate},''
  \href{http://dx.doi.org/10.1103/PhysRevLett.127.102001}{{\em Phys. Rev.
  Lett.} {\bfseries 127} no.~10, (2021) 102001},
  \href{http://arxiv.org/abs/2007.04320}{{\ttfamily arXiv:2007.04320
  [hep-ph]}}.

\bibitem{Isgur:1989vq}
N.~Isgur and M.~B. Wise, ``{Weak Decays of Heavy Mesons in the Static Quark
  Approximation},'' \href{http://dx.doi.org/10.1016/0370-2693(89)90566-2}{{\em
  Phys. Lett. B} {\bfseries 232} (1989) 113--117}.

\bibitem{Isgur:1990yhj}
N.~Isgur and M.~B. Wise, ``{Weak transition form-factors between heavy
  mesons},'' \href{http://dx.doi.org/10.1016/0370-2693(90)91219-2}{{\em Phys.
  Lett. B} {\bfseries 237} (1990) 527--530}.

\bibitem{Georgi:1990um}
H.~Georgi, ``{An Effective Field Theory for Heavy Quarks at Low-energies},''
\href{http://dx.doi.org/10.1016/0370-2693(90)91128-X}{{\em Phys. Lett.}
  {\bfseries B240} (1990) 447--450}.
%%CITATION = PHLTA,B240,447;%%.

\bibitem{Goldberger:2004jt}
W.~D. Goldberger and I.~Z. Rothstein, ``{An Effective field theory of gravity
  for extended objects},''
  \href{http://dx.doi.org/10.1103/PhysRevD.73.104029}{{\em Phys. Rev. D}
  {\bfseries 73} (2006) 104029},
  \href{http://arxiv.org/abs/hep-th/0409156}{{\ttfamily arXiv:hep-th/0409156}}.

\bibitem{Walter}
W.~Goldberger, ``{Private communications},'' 2022.

\bibitem{Politzer:1988bs}
H.~D. Politzer and M.~B. Wise, ``{Effective Field Theory Approach to Processes
  Involving Both Light and Heavy Fields},''
  \href{http://dx.doi.org/10.1016/0370-2693(88)90656-9}{{\em Phys. Lett. B}
  {\bfseries 208} (1988) 504--507}.

\bibitem{Chay:1990da}
J.~Chay, H.~Georgi, and B.~Grinstein, ``{Lepton energy distributions in heavy
  meson decays from QCD},''
  \href{http://dx.doi.org/10.1016/0370-2693(90)90916-T}{{\em Phys. Lett. B}
  {\bfseries 247} (1990) 399--405}.

\bibitem{Bigi:1992su}
I.~I.~Y. Bigi, N.~G. Uraltsev, and A.~I. Vainshtein, ``{Nonperturbative
  corrections to inclusive beauty and charm decays: QCD versus phenomenological
  models},'' \href{http://dx.doi.org/10.1016/0370-2693(92)90908-M}{{\em Phys.
  Lett. B} {\bfseries 293} (1992) 430--436},
  \href{http://arxiv.org/abs/hep-ph/9207214}{{\ttfamily arXiv:hep-ph/9207214}}.
  [Erratum: Phys.Lett.B 297, 477--477 (1992)].

\bibitem{Bigi:1993fe}
I.~I.~Y. Bigi, M.~A. Shifman, N.~G. Uraltsev, and A.~I. Vainshtein, ``{QCD
  predictions for lepton spectra in inclusive heavy flavor decays},''
  \href{http://dx.doi.org/10.1103/PhysRevLett.71.496}{{\em Phys. Rev. Lett.}
  {\bfseries 71} (1993) 496--499},
  \href{http://arxiv.org/abs/hep-ph/9304225}{{\ttfamily arXiv:hep-ph/9304225}}.

\bibitem{Manohar:1993qn}
A.~V. Manohar and M.~B. Wise, ``{Inclusive semileptonic B and polarized
  $\Lambda_b$ decays from QCD},''
  \href{http://dx.doi.org/10.1103/PhysRevD.49.1310}{{\em Phys. Rev. D}
  {\bfseries 49} (1994) 1310--1329},
  \href{http://arxiv.org/abs/hep-ph/9308246}{{\ttfamily arXiv:hep-ph/9308246}}.

\bibitem{Charles:1998dr}
J.~Charles, A.~Le~Yaouanc, L.~Oliver, O.~Pene, and J.~C. Raynal, ``{Heavy to
  light form-factors in the heavy mass to large energy limit of QCD},''
  \href{http://dx.doi.org/10.1103/PhysRevD.60.014001}{{\em Phys. Rev. D}
  {\bfseries 60} (1999) 014001},
  \href{http://arxiv.org/abs/hep-ph/9812358}{{\ttfamily arXiv:hep-ph/9812358}}.

\bibitem{Dugan:1990de}
M.~J. Dugan and B.~Grinstein, ``{QCD basis for factorization in decays of heavy
  mesons},'' \href{http://dx.doi.org/10.1016/0370-2693(91)90271-Q}{{\em Phys.
  Lett. B} {\bfseries 255} (1991) 583--588}.

\bibitem{Burdman:1998mk}
G.~Burdman, ``{Short distance coefficients and the vanishing of the lepton
  asymmetry in $B \to$ V $\ell^+$ lepton-},''
  \href{http://dx.doi.org/10.1103/PhysRevD.57.4254}{{\em Phys. Rev. D}
  {\bfseries 57} (1998) 4254--4257},
  \href{http://arxiv.org/abs/hep-ph/9710550}{{\ttfamily arXiv:hep-ph/9710550}}.

\bibitem{Beneke:1999br}
M.~Beneke, G.~Buchalla, M.~Neubert, and C.~T. Sachrajda, ``{QCD factorization
  for $B \to \pi \pi$ decays: Strong phases and CP violation in the heavy quark
  limit},'' \href{http://dx.doi.org/10.1103/PhysRevLett.83.1914}{{\em Phys.
  Rev. Lett.} {\bfseries 83} (1999) 1914--1917},
  \href{http://arxiv.org/abs/hep-ph/9905312}{{\ttfamily arXiv:hep-ph/9905312}}.

\bibitem{Beneke:2000ry}
M.~Beneke, G.~Buchalla, M.~Neubert, and C.~T. Sachrajda, ``{QCD factorization
  for exclusive, nonleptonic B meson decays: General arguments and the case of
  heavy light final states},''
  \href{http://dx.doi.org/10.1016/S0550-3213(00)00559-9}{{\em Nucl. Phys. B}
  {\bfseries 591} (2000) 313--418},
  \href{http://arxiv.org/abs/hep-ph/0006124}{{\ttfamily arXiv:hep-ph/0006124}}.

\bibitem{Bigi:1994ga}
I.~I.~Y. Bigi, M.~A. Shifman, N.~G. Uraltsev, and A.~I. Vainshtein, ``{Sum
  rules for heavy flavor transitions in the SV limit},''
  \href{http://dx.doi.org/10.1103/PhysRevD.52.196}{{\em Phys. Rev. D}
  {\bfseries 52} (1995) 196--235},
  \href{http://arxiv.org/abs/hep-ph/9405410}{{\ttfamily arXiv:hep-ph/9405410}}.

\bibitem{Czarnecki:1997sz}
A.~Czarnecki, K.~Melnikov, and N.~Uraltsev, ``{NonAbelian dipole radiation and
  the heavy quark expansion},''
  \href{http://dx.doi.org/10.1103/PhysRevLett.80.3189}{{\em Phys. Rev. Lett.}
  {\bfseries 80} (1998) 3189--3192},
  \href{http://arxiv.org/abs/hep-ph/9708372}{{\ttfamily arXiv:hep-ph/9708372}}.

\bibitem{Beneke:1998rk}
M.~Beneke, ``{A Quark mass definition adequate for threshold problems},''
  \href{http://dx.doi.org/10.1016/S0370-2693(98)00741-2}{{\em Phys. Lett. B}
  {\bfseries 434} (1998) 115--125},
  \href{http://arxiv.org/abs/hep-ph/9804241}{{\ttfamily arXiv:hep-ph/9804241}}.

\bibitem{Hoang:1998ng}
A.~H. Hoang, Z.~Ligeti, and A.~V. Manohar, ``{$B$ decay and the Upsilon
  mass},'' \href{http://dx.doi.org/10.1103/PhysRevLett.82.277}{{\em
  Phys.~Rev.~Lett.} {\bfseries 82} (1999) 277--280},
\href{http://arxiv.org/abs/hep-ph/9809423}{{\ttfamily hep-ph/9809423}}.
%%CITATION = HEP-PH/9809423;%%.

\bibitem{Hoang:1998hm}
A.~H. Hoang, Z.~Ligeti, and A.~V. Manohar, ``{$B$ decays in the upsilon
  expansion},'' \href{http://dx.doi.org/10.1103/PhysRevD.59.074017}{{\em
  Phys.~Rev.~D} {\bfseries 59} (1999) 074017},
\href{http://arxiv.org/abs/hep-ph/9811239}{{\ttfamily hep-ph/9811239}}.
%%CITATION = HEP-PH/9811239;%%.

\bibitem{Hoang:1999zc}
A.~H. Hoang and T.~Teubner, ``{Top quark pair production close to threshold:
  Top mass, width and momentum distribution},''
  \href{http://dx.doi.org/10.1103/PhysRevD.60.114027}{{\em Phys. Rev. D}
  {\bfseries 60} (1999) 114027},
  \href{http://arxiv.org/abs/hep-ph/9904468}{{\ttfamily arXiv:hep-ph/9904468}}.

\bibitem{kirill}
K.~Melnikov, ``{Private communications},'' 2022.

\bibitem{Laporta:2000dsw}
S.~Laporta, ``{High precision calculation of multiloop Feynman integrals by
  difference equations},''
  \href{http://dx.doi.org/10.1142/S0217751X00002159}{{\em Int. J. Mod. Phys. A}
  {\bfseries 15} (2000) 5087--5159},
  \href{http://arxiv.org/abs/hep-ph/0102033}{{\ttfamily arXiv:hep-ph/0102033}}.

\bibitem{Laporta:1996mq}
S.~Laporta and E.~Remiddi, ``{The Analytical value of the electron (g-2) at
  order $\alpha^3$ in QED},''
  \href{http://dx.doi.org/10.1016/0370-2693(96)00439-X}{{\em Phys. Lett. B}
  {\bfseries 379} (1996) 283--291},
  \href{http://arxiv.org/abs/hep-ph/9602417}{{\ttfamily arXiv:hep-ph/9602417}}.

\bibitem{Hill:2013hoa}
R.~J. Hill and M.~P. Solon, ``{WIMP-nucleon scattering with heavy WIMP
  effective theory},''
  \href{http://dx.doi.org/10.1103/PhysRevLett.112.211602}{{\em Phys. Rev.
  Lett.} {\bfseries 112} (2014) 211602},
  \href{http://arxiv.org/abs/1309.4092}{{\ttfamily arXiv:1309.4092 [hep-ph]}}.

\bibitem{Hill:2014yka}
R.~J. Hill and M.~P. Solon, ``{Standard Model anatomy of WIMP dark matter
  direct detection I: weak-scale matching},''
  \href{http://dx.doi.org/10.1103/PhysRevD.91.043504}{{\em Phys. Rev. D}
  {\bfseries 91} (2015) 043504},
  \href{http://arxiv.org/abs/1401.3339}{{\ttfamily arXiv:1401.3339 [hep-ph]}}.

\bibitem{Hill:2014yxa}
R.~J. Hill and M.~P. Solon, ``{Standard Model anatomy of WIMP dark matter
  direct detection II: QCD analysis and hadronic matrix elements},''
  \href{http://dx.doi.org/10.1103/PhysRevD.91.043505}{{\em Phys. Rev. D}
  {\bfseries 91} (2015) 043505},
  \href{http://arxiv.org/abs/1409.8290}{{\ttfamily arXiv:1409.8290 [hep-ph]}}.

\bibitem{Chen:2018uqz}
C.-Y. Chen, R.~J. Hill, M.~P. Solon, and A.~M. Wijangco, ``{Power Corrections
  to the Universal Heavy WIMP-Nucleon Cross Section},''
  \href{http://dx.doi.org/10.1016/j.physletb.2018.04.021}{{\em Phys. Lett. B}
  {\bfseries 781} (2018) 473--479},
  \href{http://arxiv.org/abs/1801.08551}{{\ttfamily arXiv:1801.08551
  [hep-ph]}}.

\bibitem{Hisano:2015rsa}
J.~Hisano, K.~Ishiwata, and N.~Nagata, ``{QCD Effects on Direct Detection of
  Wino Dark Matter},'' \href{http://dx.doi.org/10.1007/JHEP06(2015)097}{{\em
  JHEP} {\bfseries 06} (2015) 097},
  \href{http://arxiv.org/abs/1504.00915}{{\ttfamily arXiv:1504.00915
  [hep-ph]}}.

\bibitem{Baumgart:2017nsr}
M.~Baumgart, T.~Cohen, I.~Moult, N.~L. Rodd, T.~R. Slatyer, M.~P. Solon, I.~W.
  Stewart, and V.~Vaidya, ``{Resummed Photon Spectra for WIMP Annihilation},''
  \href{http://dx.doi.org/10.1007/JHEP03(2018)117}{{\em JHEP} {\bfseries 03}
  (2018) 117}, \href{http://arxiv.org/abs/1712.07656}{{\ttfamily
  arXiv:1712.07656 [hep-ph]}}.

\bibitem{Baumgart:2018yed}
M.~Baumgart, T.~Cohen, E.~Moulin, I.~Moult, L.~Rinchiuso, N.~L. Rodd, T.~R.
  Slatyer, I.~W. Stewart, and V.~Vaidya, ``{Precision Photon Spectra for Wino
  Annihilation},'' \href{http://dx.doi.org/10.1007/JHEP01(2019)036}{{\em JHEP}
  {\bfseries 01} (2019) 036}, \href{http://arxiv.org/abs/1808.08956}{{\ttfamily
  arXiv:1808.08956 [hep-ph]}}.

\bibitem{Bauer:2020jay}
C.~W. Bauer, N.~L. Rodd, and B.~R. Webber, ``{Dark matter spectra from the
  electroweak to the Planck scale},''
  \href{http://dx.doi.org/10.1007/JHEP06(2021)121}{{\em JHEP} {\bfseries 06}
  (2021) 121}, \href{http://arxiv.org/abs/2007.15001}{{\ttfamily
  arXiv:2007.15001 [hep-ph]}}.

\bibitem{LHCb:2021trn}
{\bfseries LHCb} Collaboration, R.~Aaij {\em et~al.}, ``{Test of lepton
  universality in beauty-quark decays},''
  \href{http://dx.doi.org/10.1038/s41567-021-01478-8}{{\em Nature Phys.}
  {\bfseries 18} no.~3, (2022) 277--282},
  \href{http://arxiv.org/abs/2103.11769}{{\ttfamily arXiv:2103.11769
  [hep-ex]}}.

\bibitem{Geng:2021nhg}
L.-S. Geng, B.~Grinstein, S.~J\"ager, S.-Y. Li, J.~Martin~Camalich, and R.-X.
  Shi, ``{Implications of new evidence for lepton-universality violation in
  $b\to s\ell^+\ell^-$ decays},''
  \href{http://dx.doi.org/10.1103/PhysRevD.104.035029}{{\em Phys. Rev. D}
  {\bfseries 104} no.~3, (2021) 035029},
  \href{http://arxiv.org/abs/2103.12738}{{\ttfamily arXiv:2103.12738
  [hep-ph]}}.

\bibitem{Altmannshofer:2021qrr}
W.~Altmannshofer and P.~Stangl, ``{New physics in rare $B$ decays after Moriond
  2021},'' \href{http://dx.doi.org/10.1140/epjc/s10052-021-09725-1}{{\em Eur.
  Phys. J. C} {\bfseries 81} no.~10, (2021) 952},
  \href{http://arxiv.org/abs/2103.13370}{{\ttfamily arXiv:2103.13370
  [hep-ph]}}.

\bibitem{Alguero:2021anc}
M.~Alguer\'o, B.~Capdevila, S.~Descotes-Genon, J.~Matias, and M.~Novoa-Brunet,
  ``{$b\rightarrow s\ell ^+\ell ^-$ global fits after $R_{K_S}$ and
  $R_{K^{*+}}$},''
  \href{http://dx.doi.org/10.1140/epjc/s10052-022-10231-1}{{\em Eur. Phys. J.
  C} {\bfseries 82} no.~4, (2022) 326},
  \href{http://arxiv.org/abs/2104.08921}{{\ttfamily arXiv:2104.08921
  [hep-ph]}}.

\bibitem{Amhis:2019ckw}
{\bfseries HFLAV} Collaboration, Y.~S. Amhis {\em et~al.}, ``{Averages of
  $b$-hadron, $c$-hadron, and $\tau$-lepton properties as of 2018},''
\href{http://arxiv.org/abs/1909.12524}{{\ttfamily arXiv:1909.12524 [hep-ex]}}.
%%CITATION = ARXIV:1909.12524;%%.

\bibitem{Bernlochner:2021vlv}
F.~U. Bernlochner, M.~F. Sevilla, D.~J. Robinson, and G.~Wormser,
  ``{Semitauonic b-hadron decays: A lepton flavor universality laboratory},''
  \href{http://dx.doi.org/10.1103/RevModPhys.94.015003}{{\em Rev. Mod. Phys.}
  {\bfseries 94} no.~1, (2022) 015003},
  \href{http://arxiv.org/abs/2101.08326}{{\ttfamily arXiv:2101.08326
  [hep-ex]}}.

\bibitem{Hashimoto:2017wqo}
S.~Hashimoto, ``{Inclusive semi-leptonic B meson decay structure functions from
  lattice QCD},'' \href{http://dx.doi.org/10.1093/ptep/ptx052}{{\em PTEP}
  {\bfseries 2017} no.~5, (2017) 053B03},
  \href{http://arxiv.org/abs/1703.01881}{{\ttfamily arXiv:1703.01881
  [hep-lat]}}.

\bibitem{Boyle:2022uba}
P.~A. Boyle {\em et~al.}, ``{A lattice QCD perspective on weak decays of b and
  c quarks Snowmass 2022 White Paper},'' in {\em {2022 Snowmass Summer Study}}.
\newblock 5, 2022.
\newblock \href{http://arxiv.org/abs/2205.15373}{{\ttfamily arXiv:2205.15373
  [hep-lat]}}.

\bibitem{Bordone:2016gaq}
M.~Bordone, G.~Isidori, and A.~Pattori, ``{On the Standard Model predictions
  for $R_K$ and $R_{K^*}$},''
  \href{http://dx.doi.org/10.1140/epjc/s10052-016-4274-7}{{\em Eur. Phys. J. C}
  {\bfseries 76} no.~8, (2016) 440},
  \href{http://arxiv.org/abs/1605.07633}{{\ttfamily arXiv:1605.07633
  [hep-ph]}}.

\bibitem{Robinson:2021cws}
D.~J. Robinson, ``{Lepton universality violation from neutral pion decays in
  RK(*) measurements},''
  \href{http://dx.doi.org/10.1103/PhysRevD.105.L031903}{{\em Phys. Rev. D}
  {\bfseries 105} no.~3, (2022) L031903},
  \href{http://arxiv.org/abs/2110.11209}{{\ttfamily arXiv:2110.11209
  [hep-ph]}}.

\bibitem{Isidori:2022bzw}
G.~Isidori, D.~Lancierini, S.~Nabeebaccus, and R.~Zwicky, ``{QED in $\bar B \to
  \bar K \ell^+\ell^-$ LFU ratios: Theory versus Experiment, a Monte Carlo
  Study},'' \href{http://arxiv.org/abs/2205.08635}{{\ttfamily arXiv:2205.08635
  [hep-ph]}}.

\bibitem{Isidori:2020acz}
G.~Isidori, S.~Nabeebaccus, and R.~Zwicky, ``{QED corrections in $
  \overline{B}\to \overline{K}{\mathrm{\ell}}^{+}{\mathrm{\ell}}^{-} $ at the
  double-differential level},''
  \href{http://dx.doi.org/10.1007/JHEP12(2020)104}{{\em JHEP} {\bfseries 12}
  (2020) 104}, \href{http://arxiv.org/abs/2009.00929}{{\ttfamily
  arXiv:2009.00929 [hep-ph]}}.

\bibitem{Barberio:1990ms}
E.~Barberio, B.~van Eijk, and Z.~Was, ``{PHOTOS: A Universal Monte Carlo for
  QED radiative corrections in decays},''
  \href{http://dx.doi.org/10.1016/0010-4655(91)90012-A}{{\em Comput. Phys.
  Commun.} {\bfseries 66} (1991) 115--128}.

\bibitem{Golonka:2005pn}
P.~Golonka and Z.~Was, ``{PHOTOS Monte Carlo: A Precision tool for QED
  corrections in $Z$ and $W$ decays},''
  \href{http://dx.doi.org/10.1140/epjc/s2005-02396-4}{{\em Eur. Phys. J. C}
  {\bfseries 45} (2006) 97--107},
  \href{http://arxiv.org/abs/hep-ph/0506026}{{\ttfamily arXiv:hep-ph/0506026}}.

\bibitem{Davidson:2010ew}
N.~Davidson, T.~Przedzinski, and Z.~Was, ``{PHOTOS interface in C++: Technical
  and Physics Documentation},''
  \href{http://dx.doi.org/10.1016/j.cpc.2015.09.013}{{\em Comput. Phys.
  Commun.} {\bfseries 199} (2016) 86--101},
  \href{http://arxiv.org/abs/1011.0937}{{\ttfamily arXiv:1011.0937 [hep-ph]}}.

\bibitem{Sachrajda}
C.~Sachrajda, ``Challenges for precision lattice flavour physics.''. Talk at
  the workshop Towards the Ultimate Precision in Flavour Physics, University of
  Warwick, Apr.\ 2018,
  \url{https://indico.cern.ch/event/694666/contributions/2916436/attachments/1633240/2604768/Sachrajda_TUPIFP.pdf}.

\bibitem{Fael:2020tow}
M.~Fael, K.~Sch\"onwald, and M.~Steinhauser, ``{Third order corrections to the
  semileptonic $b\to c$ and the muon decays},''
  \href{http://dx.doi.org/10.1103/PhysRevD.104.016003}{{\em Phys. Rev. D}
  {\bfseries 104} no.~1, (2021) 016003},
  \href{http://arxiv.org/abs/2011.13654}{{\ttfamily arXiv:2011.13654
  [hep-ph]}}.

\bibitem{Fael:2020njb}
M.~Fael, K.~Sch\"onwald, and M.~Steinhauser, ``{Relation between the
  $\overline{\mathrm{MS}}$ and the kinetic mass of heavy quarks},''
  \href{http://dx.doi.org/10.1103/PhysRevD.103.014005}{{\em Phys. Rev. D}
  {\bfseries 103} no.~1, (2021) 014005},
  \href{http://arxiv.org/abs/2011.11655}{{\ttfamily arXiv:2011.11655
  [hep-ph]}}.

\bibitem{Bordone:2021oof}
M.~Bordone, B.~Capdevila, and P.~Gambino, ``{Three loop calculations and
  inclusive $V_{cb}$},''
  \href{http://dx.doi.org/10.1016/j.physletb.2021.136679}{{\em Phys. Lett. B}
  {\bfseries 822} (2021) 136679},
  \href{http://arxiv.org/abs/2107.00604}{{\ttfamily arXiv:2107.00604
  [hep-ph]}}.

\bibitem{Mannel:2010wj}
T.~Mannel, S.~Turczyk, and N.~Uraltsev, ``{Higher Order Power Corrections in
  Inclusive B Decays},'' \href{http://dx.doi.org/10.1007/JHEP11(2010)109}{{\em
  JHEP} {\bfseries 11} (2010) 109},
  \href{http://arxiv.org/abs/1009.4622}{{\ttfamily arXiv:1009.4622 [hep-ph]}}.

\bibitem{Mannel:2021zzr}
T.~Mannel, D.~Moreno, and A.~A. Pivovarov, ``{NLO QCD corrections to inclusive
  $b \rightarrow c \ell \bar{\nu}$ decay spectra up to~$1/m_Q^3$},''
  \href{http://dx.doi.org/10.1103/PhysRevD.105.054033}{{\em Phys. Rev. D}
  {\bfseries 105} no.~5, (2022) 054033},
  \href{http://arxiv.org/abs/2112.03875}{{\ttfamily arXiv:2112.03875
  [hep-ph]}}.

\bibitem{Fael:2018vsp}
M.~Fael, T.~Mannel, and K.~Keri~Vos, ``{$V_{cb}$ determination from inclusive
  $b \to c$ decays: an alternative method},''
  \href{http://dx.doi.org/10.1007/JHEP02(2019)177}{{\em JHEP} {\bfseries 02}
  (2019) 177}, \href{http://arxiv.org/abs/1812.07472}{{\ttfamily
  arXiv:1812.07472 [hep-ph]}}.

\bibitem{Bernlochner:2022ucr}
F.~Bernlochner, M.~Fael, K.~Olschewsky, E.~Persson, R.~van Tonder, K.~K. Vos,
  and M.~Welsch, ``{First extraction of inclusive $V_{cb}$ from $q^2$
  moments},'' \href{http://arxiv.org/abs/2205.10274}{{\ttfamily
  arXiv:2205.10274 [hep-ph]}}.

\bibitem{Bobeth:2021lya}
C.~Bobeth, M.~Bordone, N.~Gubernari, M.~Jung, and D.~van Dyk, ``{Lepton-flavour
  non-universality of ${\bar{B}}\rightarrow D^*\ell {{\bar{\nu }}}$ angular
  distributions in and beyond the Standard Model},''
  \href{http://dx.doi.org/10.1140/epjc/s10052-021-09724-2}{{\em Eur. Phys. J.
  C} {\bfseries 81} no.~11, (2021) 984},
  \href{http://arxiv.org/abs/2104.02094}{{\ttfamily arXiv:2104.02094
  [hep-ph]}}.

\bibitem{Bernlochner:2020tfi}
F.~U. Bernlochner, S.~Duell, Z.~Ligeti, M.~Papucci, and D.~J. Robinson, ``{Das
  ist der HAMMER: Consistent new physics interpretations of semileptonic
  decays},'' \href{http://dx.doi.org/10.1140/epjc/s10052-020-8304-0}{{\em Eur.
  Phys. J. C} {\bfseries 80} no.~9, (2020) 883},
  \href{http://arxiv.org/abs/2002.00020}{{\ttfamily arXiv:2002.00020
  [hep-ph]}}.

\bibitem{Burgess:2021ylu}
C.~P. Burgess, S.~Hamoudou, J.~Kumar, and D.~London, ``{Beyond the standard
  model effective field theory with $B \rightarrow c \tau^- \overline{\nu}$},''
  \href{http://dx.doi.org/10.1103/PhysRevD.105.073008}{{\em Phys. Rev. D}
  {\bfseries 105} no.~7, (2022) 073008},
  \href{http://arxiv.org/abs/2111.07421}{{\ttfamily arXiv:2111.07421
  [hep-ph]}}.

\bibitem{Jenkins:2017jig}
E.~E. Jenkins, A.~V. Manohar, and P.~Stoffer, ``{Low-Energy Effective Field
  Theory below the Electroweak Scale: Operators and Matching},''
  \href{http://dx.doi.org/10.1007/JHEP03(2018)016}{{\em JHEP} {\bfseries 03}
  (2018) 016}, \href{http://arxiv.org/abs/1709.04486}{{\ttfamily
  arXiv:1709.04486 [hep-ph]}}.

\bibitem{Jenkins:2017dyc}
E.~E. Jenkins, A.~V. Manohar, and P.~Stoffer, ``{Low-Energy Effective Field
  Theory below the Electroweak Scale: Anomalous Dimensions},''
  \href{http://dx.doi.org/10.1007/JHEP01(2018)084}{{\em JHEP} {\bfseries 01}
  (2018) 084}, \href{http://arxiv.org/abs/1711.05270}{{\ttfamily
  arXiv:1711.05270 [hep-ph]}}.

\bibitem{LHCb:2014vgu}
{\bfseries LHCb} Collaboration, R.~Aaij {\em et~al.}, ``{Test of lepton
  universality using $B^{+}\rightarrow K^{+}\ell^{+}\ell^{-}$ decays},''
  \href{http://dx.doi.org/10.1103/PhysRevLett.113.151601}{{\em Phys. Rev.
  Lett.} {\bfseries 113} (2014) 151601},
  \href{http://arxiv.org/abs/1406.6482}{{\ttfamily arXiv:1406.6482 [hep-ex]}}.

\bibitem{LHCb:2021zwz}
{\bfseries LHCb} Collaboration, R.~Aaij {\em et~al.}, ``{Branching Fraction
  Measurements of the Rare $B^0_s\rightarrow\phi\mu^+\mu^-$ and
  $B^0_s\rightarrow f_2^\prime(1525)\mu^+\mu^-$ Decays},''
  \href{http://dx.doi.org/10.1103/PhysRevLett.127.151801}{{\em Phys. Rev.
  Lett.} {\bfseries 127} no.~15, (2021) 151801},
  \href{http://arxiv.org/abs/2105.14007}{{\ttfamily arXiv:2105.14007
  [hep-ex]}}.

\bibitem{Bharucha:2015bzk}
A.~Bharucha, D.~M. Straub, and R.~Zwicky, ``{$B\to V\ell^+\ell^-$ in the
  Standard Model from light-cone sum rules},''
  \href{http://dx.doi.org/10.1007/JHEP08(2016)098}{{\em JHEP} {\bfseries 08}
  (2016) 098}, \href{http://arxiv.org/abs/1503.05534}{{\ttfamily
  arXiv:1503.05534 [hep-ph]}}.

\bibitem{Horgan:2013pva}
R.~R. Horgan, Z.~Liu, S.~Meinel, and M.~Wingate, ``{Calculation of $B^0 \to
  K^{*0} \mu^+ \mu^-$ and $B_s^0 \to \phi \mu^+ \mu^-$ observables using form
  factors from lattice QCD},''
  \href{http://dx.doi.org/10.1103/PhysRevLett.112.212003}{{\em Phys. Rev.
  Lett.} {\bfseries 112} (2014) 212003},
  \href{http://arxiv.org/abs/1310.3887}{{\ttfamily arXiv:1310.3887 [hep-ph]}}.

\bibitem{Bauer:2002aj}
C.~W. Bauer, D.~Pirjol, and I.~W. Stewart, ``{Factorization and endpoint
  singularities in heavy to light decays},''
  \href{http://dx.doi.org/10.1103/PhysRevD.67.071502}{{\em Phys. Rev. D}
  {\bfseries 67} (2003) 071502},
  \href{http://arxiv.org/abs/hep-ph/0211069}{{\ttfamily arXiv:hep-ph/0211069}}.

\bibitem{Beneke:2003pa}
M.~Beneke and T.~Feldmann, ``{Factorization of heavy to light form-factors in
  soft collinear effective theory},''
  \href{http://dx.doi.org/10.1016/j.nuclphysb.2004.02.033}{{\em Nucl. Phys. B}
  {\bfseries 685} (2004) 249--296},
  \href{http://arxiv.org/abs/hep-ph/0311335}{{\ttfamily arXiv:hep-ph/0311335}}.

\bibitem{Matias:2012xw}
J.~Matias, F.~Mescia, M.~Ramon, and J.~Virto, ``{Complete Anatomy of $\bar{B}_d
  \to \bar{K}^{* 0} (\to K \pi)l^+l^-$ and its angular distribution},''
  \href{http://dx.doi.org/10.1007/JHEP04(2012)104}{{\em JHEP} {\bfseries 04}
  (2012) 104}, \href{http://arxiv.org/abs/1202.4266}{{\ttfamily arXiv:1202.4266
  [hep-ph]}}.

\bibitem{LHCb:2020gog}
{\bfseries LHCb} Collaboration, R.~Aaij {\em et~al.}, ``{Angular Analysis of
  the $B^{+}\rightarrow K^{\ast+}\mu^{+}\mu^{-}$ Decay},''
  \href{http://dx.doi.org/10.1103/PhysRevLett.126.161802}{{\em Phys. Rev.
  Lett.} {\bfseries 126} no.~16, (2021) 161802},
  \href{http://arxiv.org/abs/2012.13241}{{\ttfamily arXiv:2012.13241
  [hep-ex]}}.

\bibitem{Beneke:2009az}
M.~Beneke, G.~Buchalla, M.~Neubert, and C.~T. Sachrajda, ``{Penguins with Charm
  and Quark-Hadron Duality},''
  \href{http://dx.doi.org/10.1140/epjc/s10052-009-1028-9}{{\em Eur. Phys. J. C}
  {\bfseries 61} (2009) 439--449},
  \href{http://arxiv.org/abs/0902.4446}{{\ttfamily arXiv:0902.4446 [hep-ph]}}.

\bibitem{Lyon:2014hpa}
J.~Lyon and R.~Zwicky, ``{Resonances gone topsy turvy - the charm of QCD or new
  physics in $b \to s \ell^+ \ell^-$?},''
  \href{http://arxiv.org/abs/1406.0566}{{\ttfamily arXiv:1406.0566 [hep-ph]}}.

\bibitem{Straub:2018kue}
D.~M. Straub, ``{flavio: a Python package for flavour and precision
  phenomenology in the Standard Model and beyond},''
  \href{http://arxiv.org/abs/1810.08132}{{\ttfamily arXiv:1810.08132
  [hep-ph]}}.

\bibitem{vanDyk:2021sup}
D.~van Dyk {\em et~al.}, ``{EOS - A Software for Flavor Physics
  Phenomenology},'' \href{http://arxiv.org/abs/2111.15428}{{\ttfamily
  arXiv:2111.15428 [hep-ph]}}.

\bibitem{Arbey:2018msw}
A.~Arbey, F.~Mahmoudi, and G.~Robbins, ``{SuperIso Relic v4: A program for
  calculating dark matter and flavour physics observables in Supersymmetry},''
  \href{http://dx.doi.org/10.1016/j.cpc.2019.01.014}{{\em Comput. Phys.
  Commun.} {\bfseries 239} (2019) 238--264},
  \href{http://arxiv.org/abs/1806.11489}{{\ttfamily arXiv:1806.11489
  [hep-ph]}}.

\bibitem{ParticleDataGroup:2020ssz}
{\bfseries Particle Data Group} Collaboration, P.~A. Zyla {\em et~al.},
  ``{Review of Particle Physics},''
  \href{http://dx.doi.org/10.1093/ptep/ptaa104}{{\em PTEP} {\bfseries 2020}
  no.~8, (2020) 083C01}.

\bibitem{Frings:2015eva}
P.~Frings, U.~Nierste, and M.~Wiebusch, ``{Penguin contributions to CP phases
  in $B_{d,s}$ decays to charmonium},''
  \href{http://dx.doi.org/10.1103/PhysRevLett.115.061802}{{\em Phys. Rev.
  Lett.} {\bfseries 115} no.~6, (2015) 061802},
  \href{http://arxiv.org/abs/1503.00859}{{\ttfamily arXiv:1503.00859
  [hep-ph]}}.

\bibitem{Ligeti:2015yma}
Z.~Ligeti and D.~J. Robinson, ``{Towards more precise determinations of the
  quark mixing phase $\beta$},''
  \href{http://dx.doi.org/10.1103/PhysRevLett.115.251801}{{\em Phys. Rev.
  Lett.} {\bfseries 115} no.~25, (2015) 251801},
  \href{http://arxiv.org/abs/1507.06671}{{\ttfamily arXiv:1507.06671
  [hep-ph]}}.

\bibitem{Jung:2015yma}
M.~Jung, ``{Branching ratio measurements and isospin violation in B-meson
  decays},'' \href{http://dx.doi.org/10.1016/j.physletb.2015.12.024}{{\em Phys.
  Lett. B} {\bfseries 753} (2016) 187--190},
  \href{http://arxiv.org/abs/1510.03423}{{\ttfamily arXiv:1510.03423
  [hep-ph]}}.

\bibitem{Beneke:2005pu}
M.~Beneke, ``{Corrections to sin(2 beta) from CP asymmetries in $B^0 \to
  (\pi^0,\rho^0,\eta,\eta',\omega,\phi) K_S$ decays},''
  \href{http://dx.doi.org/10.1016/j.physletb.2005.06.045}{{\em Phys. Lett. B}
  {\bfseries 620} (2005) 143--150},
  \href{http://arxiv.org/abs/hep-ph/0505075}{{\ttfamily arXiv:hep-ph/0505075}}.

\bibitem{Williamson:2006hb}
A.~R. Williamson and J.~Zupan, ``{Two body B decays with isosinglet final
  states in SCET},'' \href{http://dx.doi.org/10.1103/PhysRevD.74.014003}{{\em
  Phys. Rev. D} {\bfseries 74} (2006) 014003},
  \href{http://arxiv.org/abs/hep-ph/0601214}{{\ttfamily arXiv:hep-ph/0601214}}.
  [Erratum: {\it Phys.\ Rev.\ D} {\bf 74}, 03901 (2006)].

\bibitem{Zupan:2007ca}
J.~Zupan, ``{Predictions for $\sin 2(\beta/\phi_1)_{\rm eff}$ in $b \to s$
  penguin dominated modes},'' {\em eConf} {\bfseries C070512} (2007) 012,
  \href{http://arxiv.org/abs/0707.1323}{{\ttfamily arXiv:0707.1323 [hep-ph]}}.

\bibitem{Grossman:2003qp}
Y.~Grossman, Z.~Ligeti, Y.~Nir, and H.~Quinn, ``{SU(3) relations and the CP
  asymmetries in B decays to $\eta' K_S$, $\phi K_S$ and $K^+ K^- K_S$},''
  \href{http://dx.doi.org/10.1103/PhysRevD.68.015004}{{\em Phys. Rev. D}
  {\bfseries 68} (2003) 015004},
  \href{http://arxiv.org/abs/hep-ph/0303171}{{\ttfamily arXiv:hep-ph/0303171}}.

\bibitem{CLEO:1997ree}
{\bfseries CLEO} Collaboration, R.~Godang {\em et~al.}, ``{Observation of
  exclusive two-body B decays to kaons and pions},''
  \href{http://dx.doi.org/10.1103/PhysRevLett.80.3456}{{\em Phys. Rev. Lett.}
  {\bfseries 80} (1998) 3456--3460},
  \href{http://arxiv.org/abs/hep-ex/9711010}{{\ttfamily arXiv:hep-ex/9711010}}.

\bibitem{Bauer:2004tj}
C.~W. Bauer, D.~Pirjol, I.~Z. Rothstein, and I.~W. Stewart, ``{$B \to M_1 M_2$:
  Factorization, charming penguins, strong phases, and polarization},''
  \href{http://dx.doi.org/10.1103/PhysRevD.70.054015}{{\em Phys. Rev. D}
  {\bfseries 70} (2004) 054015},
  \href{http://arxiv.org/abs/hep-ph/0401188}{{\ttfamily arXiv:hep-ph/0401188}}.

\bibitem{Bauer:2005kd}
C.~W. Bauer, I.~Z. Rothstein, and I.~W. Stewart, ``{SCET analysis of $B \to K
  \pi$, $B \to K \bar K$, and $B \to \pi \pi$ decays},''
  \href{http://dx.doi.org/10.1103/PhysRevD.74.034010}{{\em Phys. Rev. D}
  {\bfseries 74} (2006) 034010},
  \href{http://arxiv.org/abs/hep-ph/0510241}{{\ttfamily arXiv:hep-ph/0510241}}.

\bibitem{Beneke:2004bn}
M.~Beneke, G.~Buchalla, M.~Neubert, and C.~T. Sachrajda, ``{Comment on `$B \to
  M_1M_2$: Factorization, charming penguins, strong phases, and
  polarization'},'' \href{http://dx.doi.org/10.1103/PhysRevD.72.098501}{{\em
  Phys. Rev. D} {\bfseries 72} (2005) 098501},
  \href{http://arxiv.org/abs/hep-ph/0411171}{{\ttfamily arXiv:hep-ph/0411171}}.

\bibitem{Aaij:2019kcg}
{\bfseries LHCb} Collaboration, R.~Aaij {\em et~al.}, ``{Observation of CP
  Violation in Charm Decays},''
  \href{http://dx.doi.org/10.1103/PhysRevLett.122.211803}{{\em Phys. Rev.
  Lett.} {\bfseries 122} no.~21, (2019) 211803},
  \href{http://arxiv.org/abs/1903.08726}{{\ttfamily arXiv:1903.08726
  [hep-ex]}}.

\bibitem{LHCb:2021dcr}
{\bfseries LHCb} Collaboration, R.~Aaij {\em et~al.}, ``{Simultaneous
  determination of CKM angle $\gamma$ and charm mixing parameters},''
  \href{http://dx.doi.org/10.1007/JHEP12(2021)141}{{\em JHEP} {\bfseries 12}
  (2021) 141}, \href{http://arxiv.org/abs/2110.02350}{{\ttfamily
  arXiv:2110.02350 [hep-ex]}}.

\bibitem{Kagan:2020vri}
A.~L. Kagan and L.~Silvestrini, ``{Dispersive and absorptive $CP$ violation in
  $D^0- \overline{D^0}$ mixing},''
  \href{http://dx.doi.org/10.1103/PhysRevD.103.053008}{{\em Phys. Rev. D}
  {\bfseries 103} no.~5, (2021) 053008},
  \href{http://arxiv.org/abs/2001.07207}{{\ttfamily arXiv:2001.07207
  [hep-ph]}}.

\bibitem{Bause:2020xzj}
R.~Bause, H.~Gisbert, M.~Golz, and G.~Hiller, ``{Rare charm $c\to
  u\,\nu\bar{\nu}$ dineutrino null tests for $e^+e^-$ machines},''
  \href{http://dx.doi.org/10.1103/PhysRevD.103.015033}{{\em Phys. Rev. D}
  {\bfseries 103} no.~1, (2021) 015033},
  \href{http://arxiv.org/abs/2010.02225}{{\ttfamily arXiv:2010.02225
  [hep-ph]}}.

\bibitem{NA62:2021zjw}
{\bfseries NA62} Collaboration, E.~Cortina~Gil {\em et~al.}, ``{Measurement of
  the very rare $K^{+}\to \pi^+\nu \overline\nu$ decay},''
  \href{http://dx.doi.org/10.1007/JHEP06(2021)093}{{\em JHEP} {\bfseries 06}
  (2021) 093}, \href{http://arxiv.org/abs/2103.15389}{{\ttfamily
  arXiv:2103.15389 [hep-ex]}}.

\bibitem{DAmbrosio:2017klp}
G.~D'Ambrosio and T.~Kitahara, ``{Direct $CP$ Violation in $K \to \mu^+
  \mu^-$},'' \href{http://dx.doi.org/10.1103/PhysRevLett.119.201802}{{\em Phys.
  Rev. Lett.} {\bfseries 119} no.~20, (2017) 201802},
  \href{http://arxiv.org/abs/1707.06999}{{\ttfamily arXiv:1707.06999
  [hep-ph]}}.

\bibitem{Dery:2021mct}
A.~Dery, M.~Ghosh, Y.~Grossman, and S.~Schacht, ``{$K \to \mu^+\mu^-$ as a
  clean probe of short-distance physics},''
  \href{http://dx.doi.org/10.1007/JHEP07(2021)103}{{\em JHEP} {\bfseries 07}
  (2021) 103}, \href{http://arxiv.org/abs/2104.06427}{{\ttfamily
  arXiv:2104.06427 [hep-ph]}}.

\bibitem{Dery:2021vql}
A.~Dery and M.~Ghosh, ``{$K \to \mu^+\mu^-$ beyond the standard model},''
  \href{http://dx.doi.org/10.1007/JHEP03(2022)048}{{\em JHEP} {\bfseries 03}
  (2022) 048}, \href{http://arxiv.org/abs/2112.05801}{{\ttfamily
  arXiv:2112.05801 [hep-ph]}}.

\bibitem{Goudzovski:2022vbt}
E.~Goudzovski {\em et~al.}, ``{New Physics Searches at Kaon and Hyperon
  Factories},'' \href{http://arxiv.org/abs/2201.07805}{{\ttfamily
  arXiv:2201.07805 [hep-ph]}}.

\end{thebibliography}\endgroup

\end{document}